\documentclass[showpacs,preprintnumbers,amsmath,amssymb,superscriptaddress,pra,twocolumn]{revtex4}

\include{amsmath}
\usepackage{graphicx}
\usepackage{subfigure}
\usepackage{amsmath}
\usepackage{amsfonts}

\newcommand{\M}{\mathrm{M}}

\newcommand{\C}{\mathrm{C}}

\newcommand{\LL}{\mathrm{L}}
\newcommand{\RR}{\mathrm{R}}
\newcommand{\Hhat}{\hat{H}}
\newcommand{\ccr}{\hat{c}^\dagger}
\newcommand{\cde}{\hat{c}}
\newcommand{\acr}{\hat{a}^\dagger}
\newcommand{\ade}{\hat{a}}

\newcommand{\zhat}{\hat{z}}
\newcommand{\etahat}{\hat{\eta}}
\newcommand{\zetahat}{\hat{\zeta}}
\newcommand{\xihat}{\hat{\xi}}
\newcommand{\dhat}{\hat{d}}
\newcommand{\Xhat}{\hat{X}}
\newcommand{\Yhat}{\hat{Y}}
\newcommand{\du}{d}
\newcommand{\im}{i}
\newcommand{\ex}{e}

\begin{document}

\title[Observability of radiation pressure shot noise in optomechanical systems]
{Observability of radiation pressure shot noise in optomechanical systems}
\author{K. B{\o}rkje}
\author{A. Nunnenkamp}
\author{B. M. Zwickl}
\author{C. Yang}
\author{J. G. E. Harris}
\author{S. M. Girvin}
\affiliation{Departments of Physics and Applied Physics, Yale University, New Haven, Connecticut 06520, USA}

\date{Received \today}
\begin{abstract}
 We present a theoretical study of an experiment designed to detect radiation pressure shot noise in an optomechanical system. Our model consists of a coherently driven optical cavity mode that is coupled to a mechanical oscillator. We examine the cross-correlation between two quadratures of the output field from the cavity. We determine under which circumstances radiation pressure shot noise can be detected by a measurement of this cross-correlation. This is done in the general case of nonzero detuning between the frequency of the drive and the cavity resonance frequency. We study the qualitative features of the different contributions to the cross-correlator and provide quantitative figures of merit for the relative importance of the radiation pressure shot noise contribution to other contributions. We also propose a modified setup of this experiment relevant to the ``membrane-in-the-middle'' geometry, which potentially can avoid the problems of static bistability and classical noise in the drive.
\end{abstract}
\maketitle

\section{Introduction}

An optomechanical system is characterized by an interaction between light and mechanical motion \cite{Kippenberg2008Science,Marquardt2009Physics}. The recent interest in such systems was initiated by the efforts to detect gravitational waves \cite{Abromovici1992Science,Barish1999PhysToday}. The field has now taken on a life of its own, especially since the observation of quantum effects in mechanical systems is nearly within experimental reach. Besides the potential for technological innovation, the possibility to study mechanical systems in the quantum regime \cite{Kippenberg2008Science,Marquardt2009Physics} could provide insight into the fundamentals of quantum mechanics \cite{Marshall2003PRL,Kleckner2008NJP}.

The canonical optomechanical system consists of an optical cavity where one of the end mirrors is free to move \cite{Fabre1994PRA,Jacobs1994PRA}. When light from a laser enters the cavity, the light exerts a force on the movable mirror. As the mirror moves, the cavity length changes, altering the resonance frequency of the cavity and thus the optical intensity in the cavity. This in turn changes the optical force on the mirror, such that the optical and mechanical dynamics are coupled. Experimental studies of this system \cite{Metzger2004Nature,Gigan2006Nature,Arcizet2006Nature,Kleckner2006Nature,Hertzberg2010NatPhys,Groblacher2009Nature} are now getting close to observing quantum effects due to the optomechanical coupling. 

Other experimental realizations of coupled optical and mechanical degrees of freedom have emerged over the last decade. One realization includes placing a delicate membrane inside an ordinary optical cavity with fixed end mirrors \cite{Thompson2008Nature,Jayich2008NJP}. This has the advantage of not having to combine the flexibility needed for the mechanical oscillator with the rigidity of a high finesse cavity mirror. We will refer to this setup as the membrane-in-the-middle geometry. Other examples include mechanical breathing modes in toroidal microresonators \cite{Schliesser2006PRL,Schliesser2008NJP}, nanobeams coupled to microwave resonators in superconducting circuits \cite{Regal2008NatPhys,Rocheleau2009Nature}, optical forces on free-standing waveguides \cite{Li2008Nature,Eichenfield2009Nature}, and coupling to collective movement of cold atoms in an optical lattice \cite{Murch2008NatPhys,Brennecke2008Science}.

One of the major goals in the field of optomechanics is the observation of radiation pressure shot noise (RPSN). This is the radiation pressure fluctuations experienced by the mechanical oscillator due to photon number fluctuations. Equivalently, RPSN is the quantum back-action of an optical displacement measurement \cite{Caves1980PRL}. One approach to observe RPSN would be to measure a correlation between photon number fluctuations (shot noise) in the optical field and the position fluctuations of the mechanical oscillator. The main obstacle to observing this correlation is that the shot noise induced fluctuations of the mechanical oscillator are typically very small compared to thermal fluctuations associated with the mechanical damping. To detect such a correlation in the movable-mirror geometry, Heidmann {\it et al.}~\cite{Heidmann1997ApplPhysB} proposed using two optical beams, one strong signal beam and one weak meter beam. The position fluctuations of the mechanical oscillator induced by the signal beam would be detected with the meter beam. If the beams are exactly at the cavity mean resonance frequency, a correlation between the fluctuations in the signal beam intensity and the phase quadrature in the meter beam will be due to RPSN alone. 

The idea presented by Heidmann {\it et al.}~\cite{Heidmann1997ApplPhysB}  has been tested in the classical domain \cite{Verlot2009PRL}. It was shown experimentally that this scheme works when imposing additional classical noise in the beam, but correlations due to the smaller quantum shot noise have not yet been observed. A more indirect observation of RPSN was achieved in a cold atom experiment \cite{Murch2008NatPhys} through the measurement of RPSN-induced heating of the atomic ensemble.

In this article, we report theoretical studies of an experiment similar to the one proposed in Ref.~\cite{Heidmann1997ApplPhysB}. We consider a two-sided cavity, where the cavity mode is driven by one rather than two beams and coupled to a mechanical oscillator. We examine the cross-correlation between two quadratures of the cavity output field, studying the general case where the beam is not at resonance with the cavity. The qualitative features of the various contributions to the cross-correlation are explored, and we determine under which circumstances the RPSN contribution can dominate over the contribution from thermal noise. This is quantified by simple figures of merit that can be used to determine the parameter values needed in such an experiment. Furthermore, we compare the RPSN contribution to the unwanted contribution from classical noise in the laser. Finally, we also propose a modified setup for the membrane-in-the-middle geometry in which two optical modes couple to the mechanical oscillator with opposite signs. This has the potential to overcome the problem of static mechanical bistability \cite{Dorsel1983PRL,Jayich2008NJP}, as well as to significantly reduce the contribution from classical laser noise.

The article is organized as follows: In Section \ref{sec:Model}, we present the model and briefly discuss its properties. Section \ref{sec:detection} gives details on the detection scheme, presents the various contributions to the cross-correlation, and shows how to minimize the thermal noise contribution. The comparison of the quantum and thermal contributions is given in Section \ref{sec:quanttherm}, whereas Section \ref{sec:quantclass} contains the comparison to the contribution from classical laser noise. In Section \ref{sec:twomodes}, the new two-mode setup is presented. Our conclusions are presented in Section \ref{sec:conclusion}, and mathematical details can be found in Appendix \ref{app:genExp} and \ref{app:TwoModes}.

\section{Model}
\label{sec:Model}

We consider one optical mode in a two-sided cavity coupled linearly to the position of a mechanical oscillator. The Hamiltonian is
\begin{equation}
 \label{eq:hamiltonian}
  \Hhat = \hbar \omega_\M \ccr \cde + \hbar \left( \omega_\C + A \zhat \right) \left(\acr \ade - \langle \acr \ade \rangle \right) + \hat{H}_\kappa + \hat{H}_{\gamma}
\end{equation}
where $\cde$($\ade$) is the annihilation operator for the mechanical(optical) oscillator with frequency $\omega_\M$($\omega_\C$). The dimensionless position operator of the mechanical oscillator is $\zhat = \cde + \ccr$ and $A$ is a coupling constant. $\Hhat_\kappa$ and $\Hhat_{\gamma}$ contain the couplings to an optical and mechanical bath, respectively, and describe drive and/or decay of the oscillators. In the membrane-in-the-middle setup \cite{Thompson2008Nature,Jayich2008NJP}, when the cavity frequency depends linearly on membrane position, this model is valid in the limit of small membrane reflectivities.

\begin{figure}[htbp]
\begin{center} \includegraphics[width=.46\textwidth]{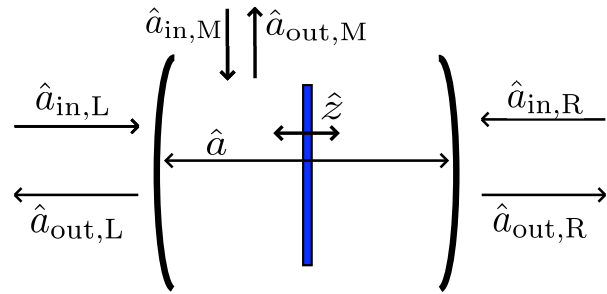}
\caption{(color online). Schematic overview of the model. The optical cavity mode $\ade$ is coupled to three input modes and to the position fluctuations $\zhat$ of the mechanical oscillator, represented in this case by a membrane in the middle. We imagine the cavity being coherently driven by a laser on the left hand side (see Figure \ref{fig:detection}), such that the input mode $\ade_{\mathrm{in,L}}$ is the sum of a mean amplitude, classical laser noise, and quantum noise. The mode $\ade_{\mathrm{in,R}}$ simply represents quantum noise entering the cavity from the right hand side, whereas $\ade_{\mathrm{in,M}}$ represents quantum noise associated with other types of decay.}
\label{fig:inputoutput}
\end{center}
\end{figure}

In the Markov approximation, input-output theory \cite{Walls2008Book,Clerk2008RMP} gives the quantum Langevin equations
\begin{eqnarray}
  \label{eq:quantLangevin}
\dot{\ade} & = & - \left(\frac{\kappa}{2} + \im \omega_\C \right) \ade - \im A \zhat \ade + \sqrt{\kappa_\LL} \ade_{\mathrm{in},\LL}  \\ & + & \sqrt{\kappa_\RR} \ade_{\mathrm{in},\RR} + \sqrt{\kappa_\M} \ade_{\mathrm{in},\M} \notag \\
  \dot{\cde} & = & - \left(\frac{\gamma}{2} + \im \omega_\M \right) \cde - \im A \left(\acr \ade -  \langle \acr \ade \rangle \right) + \sqrt{\gamma} \etahat \ . \nonumber
\end{eqnarray}
Here, $\ade_{\mathrm{in},\LL}$($\ade_{\mathrm{in},\RR}$) is the input mode at the left(right) hand side of the cavity. See Figure \ref{fig:inputoutput} for a schematic picture of the input-output modes. The decay rate due to a finite transmission of the end mirrors is characterized by $\kappa_\LL$ and $\kappa_\RR$.  We have included a third decay channel of strength $\kappa_\M$, which can describe optical loss, e.g., due to scattering of photons into other optical modes or absorption in the end mirrors or membrane. The input mode $\ade_{\mathrm{in},\M}$ is the associated quantum noise operator. The total decay rate of the cavity mode is the sum of the contribution from each port, {\it i.e.}~$\kappa = \kappa_\LL + \kappa_\RR + \kappa_\M$. The optical output modes are given by
\begin{eqnarray}
  \label{eq:input-output}
  \ade_{\mathrm{out},i}(t)  = \sqrt{\kappa_i} \ade(t) - \ade_{\mathrm{in},i}(t) \ , \ i = \mathrm{L, R, M} \ .
\end{eqnarray}

The mechanical input mode $\etahat$ describes noise from the mechanical bath. In the case of high mechanical quality factor $\omega_\M/\gamma$, we can assume
\begin{eqnarray}
  \label{eq:etaProperties}
  \langle \etahat(t) \etahat^\dagger(t') \rangle & = & (n_\mathrm{th} + 1) \, \delta(t-t') \\
 \langle \etahat^\dagger(t) \etahat(t') \rangle & = & n_\mathrm{th}  \, \delta(t-t') , \nonumber
\end{eqnarray}
where $n_\mathrm{th}$ is the mean number of phonons in the absence of optomechanical coupling, determined by the temperature of the mechanical bath \footnote{The thermal phonon occupation number is $n_\mathrm{th} = [\ex^{\hbar \omega_\M/(k_\mathrm{B} T)} - 1]^{-1}$, where $T$ is the temperature of the mechanical bath.}.

We will assume that the cavity mode is driven at the frequency $\omega_\mathrm{D}$ from the left hand side of the cavity. See Figure \ref{fig:detection} for a schematic setup. Thus, we can write
\begin{eqnarray}
  \label{eq:inputModes}
  \ade_{\mathrm{in},\LL}(t) & = & \ex^{-\im \omega_\mathrm{D} t} \left(\bar{a}_{\mathrm{in}}(t) + \xihat_\LL(t) \right) \\
  \ade_{\mathrm{in},i}(t) & = & \ex^{-\im \omega_\mathrm{D} t} \xihat_i(t) \ , \ i = \mathrm{R, M} \ . \nonumber
\end{eqnarray}
We write the coherent state amplitude as $\bar{a}_{\mathrm{in}}(t) = \bar{a}_0 + \delta x(t) + \im \delta y(t)$, where $\bar{a}_0$ is defined to be constant and real, such that the real functions $\delta x(t)$ and $\delta y(t)$ describe classical amplitude and phase fluctuations of the drive, respectively, e.g., arising from technical laser noise. The noise operators $\xihat_i$ satisfy
 \begin{eqnarray}
   \label{eq:xiProperties}
\langle \xihat_i (t) \xihat^\dagger_i (t') \rangle & = & (n_\C + 1) \, \delta(t-t') \\
 \langle \xihat^\dagger_i (t) \xihat_i (t') \rangle & = & n_\C  \, \delta(t-t') \nonumber
\end{eqnarray}
to a good approximation, where $n_\C$ is the thermal occupation number at the cavity resonance frequency \cite{Clerk2008RMP}. We will make the experimentally relevant assumption that $\hbar \omega_\C \gg k_\mathrm{B} T$ and set $n_\C$ to zero. We will also assume that the classical amplitude and phase noise is white, since we are only interested in the noise around the mechanical frequency $\omega_\M$. We denote the strength of the classical noises by $C_X$ and $C_Y$:
\begin{eqnarray}
  \label{eq:classProperties}
  \langle \delta x (t) \delta x(t') \rangle & = & C_X \, \delta(t-t') \\
 \langle \delta y (t) \delta y(t') \rangle & = & C_Y   \, \delta(t-t')  \ . \nonumber
\end{eqnarray}
$C_X$ and $C_Y$ are typically both functions of the laser power. We assume the amplitude and phase noise to be uncorrelated.

In the frame rotating at the optical drive frequency, we write the cavity amplitude as the sum of its mean value and a fluctuating part, $\ade(t) = \ex^{-\im \omega_\mathrm{D} t} (\bar{a} + \dhat(t))$. When we neglect the small terms $\dhat^\dagger \dhat$ and $\dhat \zhat$, the linearized equations of motion become
\begin{eqnarray}
  \label{eq:linearEOM}
  \dot{\dhat} & = & - \left(\frac{\kappa}{2} - \im \Delta \right) \dhat - \im \alpha \zhat + \sqrt{\kappa_\LL} \left(\delta x + \im \delta y + \xihat_\LL\right) \\
& + & \sqrt{\kappa_\RR} \xihat_\RR + \sqrt{\kappa_\M} \xihat_\M \notag \\
  \dot{\cde} & = & - \left(\frac{\gamma}{2} + \im \omega_\M\right)\cde  - \im \left(\alpha^\ast \dhat + \alpha \dhat^\dagger\right)+ \sqrt{\gamma} \etahat \ . \nonumber
\end{eqnarray}
Here, the detuning $\Delta = \omega_\mathrm{D} - \omega_\C$, the coupling $\alpha \equiv A \bar{a}$, and the mean cavity amplitude $\bar{a} = \langle \ade \rangle = \sqrt{\kappa_\LL} \bar{a}_0/(\kappa/2 - \im \Delta)$. Note that at nonzero detuning $\Delta$, the mean cavity amplitude $\bar{a}$ is phase shifted relative to the incident amplitude by a phase $\phi$ given by
\begin{equation}
  \label{eq:phidef}
   \phi = \arctan \frac{2 \Delta}{\kappa} \ .
\end{equation}

The solutions to the equations of motion are given in Equation (\ref{eq:zSolution}) of Appendix \ref{app:genExp}. They can be expressed in terms of the mechanical susceptibility \footnote{We define the Fourier transform as $\hat{f}[\omega] = \int_{-\infty}^{\infty} \du t \, \hat{f}(t) \ex^{\im \omega t}$ and $\hat{f}^\dagger[\omega] = \int_{-\infty}^{\infty} \du t \, \hat{f}^\dagger(t) \ex^{\im \omega t}$, such that $(\hat{f}^\dagger[-\omega])^\dagger = \hat{f}[\omega]$. The inverse transform is $\hat{f}^{(\dagger)}(t) = \frac{1}{2\pi} \int_{-\infty}^\infty d \omega \hat{f}^{(\dagger)}[\omega] \ex^{-\im \omega t}$.\label{foot:Fourier}} $\chi_\M[\omega] = [\gamma/2 - \im \left(\omega - \omega_\M \right)]^{-1}$, the cavity susceptibility $\chi_\C[\omega] = \left[\kappa/2 - \im (\omega + \Delta)\right]^{-1}$, and the optomechanical ``self-energy''
\begin{equation}
  \label{eq:Sigmadef}
  \Sigma[\omega] = -\im |\alpha|^2 \left(\chi_\C[\omega] - \chi_\C^\ast[-\omega] \right) \ .
\end{equation}
We will let $\gamma_\mathrm{opt}$ denote the additional damping of the mechanical oscillator due to the optomechanical interaction. In the weak-coupling limit where $|\gamma_\mathrm{opt}| \ll \kappa,\omega_\M$, the mechanical frequency and damping rate are shifted by $\delta \omega_\M = \mathrm{Re} \, \Sigma[\omega_\M]$ and $\gamma_\mathrm{opt} = -2 \, \mathrm{Im} \, \Sigma[\omega_\M]$, respectively \cite{Marquardt2007PRL}. We will assume to be in this limit and let
\begin{eqnarray}
  \label{eq:shiftedOmegaGamma}
  \tilde{\omega}_\M & = & \omega_\M + \delta \omega_\M \\
  \tilde{\gamma} & = & \gamma + \gamma_\mathrm{opt} \nonumber
\end{eqnarray}
denote the effective mechanical frequency and damping rate. The effective mean phonon number becomes
\begin{equation}
  \label{eq:nMeff}
  n_\M = \frac{\gamma \, n_\mathrm{th} + \gamma_\mathrm{opt} \, n_\mathrm{opt}}{\tilde{\gamma}} ,
\end{equation}
where $n_\mathrm{opt}$ is given in Equation (\ref{eq:noptdef}). In the absence of classical laser noise, $n_\mathrm{opt} = - \left(4 \omega_\M \Delta |\chi_C[\omega_\M]|^2\right)^{-1}$ is a measure of the effective temperature of the RPSN \cite{Marquardt2007PRL,Clerk2008RMP}. At a sufficiently large positive $\Delta$, the effective damping $\tilde{\gamma}$ becomes negative and the mechanical oscillator becomes unstable \footnote{In this case, nonlinear effects become important and the oscillator settles into periodic self-sustained oscillations. See Ref.~\onlinecite{Metzger2008PRL} and references therein.}. At negative $\Delta$, the optical damping $\gamma_\mathrm{opt}$ is positive, leading to cooling of the mechanical motion.

\section{Detecting radiation pressure shot noise}
\label{sec:detection}

As proposed by Heidmann {\it et al.}~\cite{Heidmann1997ApplPhysB}, RPSN on the mechanical oscillator can be detected by a cross-correlation measurement of the outgoing beams from the cavity. We consider a similar detection scheme here. The general idea is presented in Section \ref{sec:GenScheme}, and Section \ref{sec:Motivation} provides a motivation. In Section \ref{sec:Thermal}, we show why mechanical thermal noise poses a problem and how it can be avoided. In Section \ref{sec:GenExpressions}, we present a modified version of the experiment, and give general expressions for the various contributions to the cross-correlation function.

\subsection{General scheme}
\label{sec:GenScheme}

\begin{figure}[htbp]
\begin{center} \includegraphics[width=.46\textwidth]{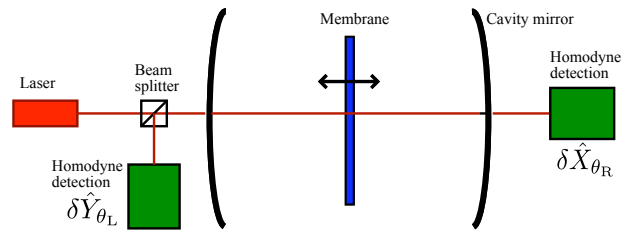}
  \caption{(color online). Schematic overview of the experiment considered. The cavity mode is coherently driven by a laser from the left hand side. One quadrature $\Xhat_{\theta_R}$ of the transmitted beam and one quadrature $\Yhat_{\theta_L}$ of the reflected beam are measured through homodyne detection. The RPSN felt by the mechanical oscillator can be detected through the cross-correlation of the fluctuations in the two quadratures.}
\label{fig:detection}
\end{center}
\end{figure}

Figure \ref{fig:detection} shows a simplified schematic of the most general setup we consider.

We assume that the quadrature
\begin{equation}
  \label{eq:Xquad0}
  \Xhat_{\theta_\RR}(t) = \ex^{\im (\omega_\mathrm{D} t -\theta_\RR)} \hat{a}_{\mathrm{out},\RR}(t) + \ex^{-\im (\omega_\mathrm{D} t -\theta_\RR)} \hat{a}^\dagger_{\mathrm{out},\RR}(t)
\end{equation}
of the transmitted beam is measured through homodyne detection, with some as yet unspecified local oscillator phase $\theta_\RR$. The fluctuations around the mean value of $\Xhat_{\theta_\RR}(t)$ can be written
\begin{equation}
  \label{eq:Xquad}
  \delta \Xhat_{\theta_\RR}(t) = \ex^{-\im \theta_\RR} \dhat_{\mathrm{out},\RR}(t)  + \ex^{\im \theta_\RR} \dhat^\dagger_{\mathrm{out},\RR}(t) \ ,
\end{equation}
where $\dhat_{\mathrm{out},\RR}(t) = \sqrt{\kappa_\RR} \dhat(t) - \xihat_\RR(t)$. Similarly, on the left hand side of the cavity, we assume that the quadrature fluctuation 
\begin{equation}
  \label{eq:Yquad}
  \delta \Yhat_{\theta_\LL}(t) = \ex^{-\im \theta_\LL} \dhat_{\mathrm{out,\LL}} (t) + \ex^{\im \theta_\LL} \dhat^\dagger_{\mathrm{out,\LL}} (t) \ ,
\end{equation}
is measured through homodyne detection. Here, $\dhat_{\mathrm{out,\LL}}(t) = \sqrt{\kappa_\LL} \dhat(t) - (\delta x(t) + \im \delta y(t) + \xihat_\LL(t))$.

For the detection of RPSN, we examine the Fourier transform of the cross-correlation of the two quadrature fluctuations $\delta \Xhat_{\theta_\RR}$ and $\delta \Yhat_{\theta_\LL}$ defined above. Since they in general do not commute, it is important to note that a measurement will correspond to the symmetrized cross-correlation. Hence, we study the function
\begin{eqnarray}
  \label{eq:Sdef}
  S[\omega] & = & \frac{1}{2}\int_{-\infty}^{\infty} \du t \, \ex^{\im \omega t} \langle \{\delta \Xhat_{\theta_\RR}(t) , \delta \Yhat_{\theta_\LL}(0)\} \rangle \\
 & = & \frac{1}{4\pi} \int_{-\infty}^{\infty} \du \omega' \, \langle \{\delta \Xhat_{\theta_\RR}[\omega] , \delta \Yhat_{\theta_\LL}[\omega']\} \rangle \ , \notag
\end{eqnarray}
where the brackets denote the anticommutator. In practice, $S[\omega]$ can easily be measured by multiplying the complex Fourier-transforms of time traces of $\delta \Xhat_{\theta_\RR}(t)$ and $\delta \Yhat_{\theta_\LL}(t)$ \footnote{From the cross-correlation theorem, the measurement of $S[\omega]$ is given by the product $\delta \Xhat_{\theta_\RR , \tau} [\omega] \delta \Yhat^\ast_{\theta_\LL , \tau} [\omega]$, where the windowed Fourier transforms are based on a sampling time $\tau$. Another possibility is of course to calculate the correlation function in real time and then determine its Fourier transform.}. We choose not to normalize $S[\omega]$ by the auto-correlators as in Ref.~\cite{Heidmann1997ApplPhysB}, since we are not only looking for a nonzero value, but are also interested in the qualitative frequency dependence, which is complicated by normalization. Also, comparing the size of the signal for different parameter values can be of interest when it comes to the problem of technical noise.


\subsection{Motivation}
\label{sec:Motivation}

The interaction part of the Hamiltonian (\ref{eq:hamiltonian}) can in the linearized case be written $\Hhat_\mathrm{int} = - l_0 \hat{z} \hat{F}$, where the optical force operator is
\begin{equation}
  \label{eq:opticalforce}
  \hat{F} = - \frac{ \hbar |\alpha|}{l_0} (\ex^{-\im \phi} \dhat + \ex^{\im \phi} \dhat^\dagger) \ ,
\end{equation}
and $l_0$ is the size of the zero point fluctuations of the oscillator. We now briefly motivate how the detection of optomechanical correlations, i.e., correlations between the optical force $\hat{F}$ and the mechanical oscillator position $\hat{z}$, can be inferred from the measurement of $S[\omega]$. 

We consider a simplified situation where the laser is on resonance with the cavity ($\Delta = 0$). In addition, we neglect the quantum vacuum noise $\xihat_i$ and the classical phase noise $\delta y(t)$. Let us at first assume that the classical amplitude noise $\delta x$ is simply a periodic signal at frequency $\omega_\mathrm{N}$, i.e., $\delta x(t) \sim \cos(\omega_\mathrm{N} t)$. Furthermore, we let $\theta_\RR = 0$, such that $\Xhat_{\theta_\RR}$ is the intensity quadrature on the right hand side, and $\theta_\LL = \pi/2$, such that $\Yhat_{\theta_\LL}$ is the phase quadrature on the left hand side. Finally, we assume that the cavity decay rate is much larger than the oscillation frequency of $\delta x$ and the resonance frequency of the mechanical oscillator, i.e., $\kappa \gg \omega_\mathrm{N}, \omega_M$. We drop hats on operators in the discussion below, since all quantum effects are neglected.

In this simplified case, the quadrature $\delta X_{0} $ is simply proportional to the optical force $F(t)$, which again is proportional to $\delta x(t)$. On the other hand, the reflected phase quadrature is proportional to the position of the mechanical oscillator, $\delta Y_{\pi/2} \sim z(t)$. The cross-correlation between the two quadratures becomes
\begin{equation}
  \label{eq:CrosscorrResonance}
  \langle \{\delta X_0(t) , \delta Y_{\pi/2}(0)\} \rangle \sim \langle F(t)  z(0) \rangle \ .
\end{equation}
This means that the function $S[\omega]$ is the Fourier transform of the cross-correlation between the optical force and the position of the oscillator. 

We can take this simple analysis one step further and consider what kind of qualitative behaviour we can expect for the function $S[\omega]$. The force is simply proportional to the signal $\delta x$, such that $F(t) \sim \cos(\omega_\mathrm{N} t)$. If we neglect the mechanical bath and assume $\gamma, |\omega_\mathrm{N}-\omega_\M| \ll \omega_\M$, the position of the mechanical oscillator is 
\begin{equation}
  \label{eq:ZExample}
  z(t) \sim  |\chi_\M[\omega_\mathrm{N}]| \cos\big(\omega_\mathrm{N} t - \lambda(\omega_\mathrm{N})\big) \ .
\end{equation}
This is phase shifted relative to the force, where $\lambda(\omega_\mathrm{N})$ is given by
\begin{equation}
  \label{eq:phaseForced}
  \lambda(\omega_\mathrm{N}) = \arctan \frac{\gamma/2}{\omega_\M  - \omega_\mathrm{N}} \ .
\end{equation}
As expected for a damped and driven harmonic oscillator, $\lambda \approx 0$ for frequencies $\omega_\M - \omega_\mathrm{N} \gg \gamma$, $\lambda = \pi/2$ for $\omega_\mathrm{N} = \omega_\M$, and $\lambda \approx \pi$ for $\omega_\mathrm{N} - \omega_\M \gg \gamma$. 

Interpreting the expectation value in Equation (\ref{eq:Sdef}) as a time average, one arrives at $S[\omega] = \tilde{S}[\omega] + \tilde{S}^\ast[-\omega]$, with
\begin{equation}
  \label{eq:SExample}
  \tilde{S}[\omega] \sim \frac{\omega_\M - \omega_\mathrm{N} + \im \gamma/2}{\left(\gamma/2\right)^2 + \left(\omega_\mathrm{N} - \omega_\M\right)^2} \delta(\omega - \omega_\mathrm{N}) \ .
\end{equation}
Remember that we assumed $\delta x(t) \sim \cos(\omega_\mathrm{N} t)$. In the relevant situation where $\delta x(t)$ represents white noise, the force is driving the oscillator at all frequencies simultaneously. The real part of the Fourier transformed cross-correlation $S[\omega]$ will then be proportional to $|\chi_\M[\omega]| \cos \lambda(\omega)$, which has a sign change at the mechanical frequency $\omega_\M$. The imaginary part will be proportional to $|\chi_\M[\omega]| \sin \lambda(\omega)$, which is a Lorentzian centered at $\omega_\M$.

The above analysis gives an idea of how optomechanical correlations manifest themselves in the function $S[\omega]$. However, we only considered a very simplified case, and the general situation is more complicated. In addition, we only took classical noise into account, but for the purpose of detecting RPSN, we are looking for correlations between {\it quantum} optical noise and oscillator position. It turns out that in the special case we analyzed above, the result is valid also for quantum noise. However, we will see that optomechanical correlations of quantum origin can actually be distinguished from their classical counterparts under certain conditions.

In general, to affirm that photon shot noise in the cavity is influencing the mechanical oscillator fluctuations, it is simply enough to prove that correlations exist between the position operator $\zhat$ and the vacuum noise operators $\xihat_i$ which are the sources of the shot noise. We will see that, in the general case, $S[\omega]$ contains such correlation functions, and we will determine under which circumstances additional terms can be neglected.

\subsection{The problem of thermal noise}
\label{sec:Thermal}

In the simplified example in the previous section, the transmitted intensity quadrature was independent of the oscillator position. This is always the case when the detuning $\Delta$ is exactly zero, as noted by Heidmann {\it et al.} \cite{Heidmann1997ApplPhysB}. In general however, both quadratures $\delta \Xhat_{\theta_\RR}$ and $\delta \Yhat_{\theta_\LL}$ depend on the oscillator position $\zhat$. This gives rise to a term in $S[\omega]$, denoted $S_\mathrm{z,z}[\omega]$ below, that is proportional to the spectral density of the mechanical oscillator, which is typically dominated by thermal noise due to the mechanical bath. If this term is much larger than the terms originating from optomechanical correlations, RPSN detection by this method becomes difficult. However, we now show that the thermal noise contribution can be made to vanish not only for zero detuning, but for any detuning $\Delta$. This will permit the observation of the optomechanical correlations.

From Equation (\ref{eq:linearEOM}), we see that the cavity field fluctuations $\hat{d}$ depend on the position operator $\zhat$, the quantum vacuum noise $\xihat_i$, and the classical laser noise $\delta x, \delta y$. Since the equations of motion are linear, we can focus on the part that depends on $\zhat$ and ignore the other terms. The cavity field fluctuations then become
\begin{equation}
  \label{eq:dzdependence}
  \hat{d}(t) = -\im \alpha \int_{-\infty}^t d \tau \, \ex^{-(\kappa/2 - \im \Delta)(t-\tau)} \zhat(\tau)
\end{equation}
Let us first consider the bad cavity limit $\kappa \gg \omega_\M$, where the cavity field follows the motion of the oscillator adiabatically. In this limit, a general quadrature of the cavity field becomes
\begin{equation}
  \label{eq:cavityQuadBadLimit}
  \ex^{-\im \theta} \hat{d}(t) + \ex^{\im \theta} \hat{d}^\dagger(t) = - \frac{2 |\alpha| \sin(\theta - 2\phi)}{\sqrt{(\kappa/2)^2 + \Delta^2}} \zhat(t) \ .
\end{equation}
This shows that in the bad cavity limit, choosing $\theta_\RR$($\theta_\LL$) to be $2 \phi$ makes $\Xhat_{\theta_\RR}$($\delta \Yhat_{\theta_\LL}$) independent of the position $\zhat$, such that the thermal noise contribution in $S[\omega]$ vanishes. This also connects smoothly with the resonant case $\Delta = 0$, where the intensity quadrature ($\theta = \phi = 0$) becomes independent of $\zhat$.

If the experiment is not in the bad cavity limit, the picture is more complicated, but it is still possible to avoid the thermal noise. We show this by first assuming that the motion of the oscillator is given simply by $z(t) \sim \cos(\omega_\mathrm{N} t)$, i.e., a harmonic oscillation at frequency $\omega_\mathrm{N}$. Again, we drop operator hats as everything is considered to be classical. The cavity quadrature given by $\theta$ becomes
\begin{eqnarray}
  \label{eq:cavityQuadGeneral}
  \ex^{-\im \theta} d(t) + \ex^{\im \theta} d^\dagger(t) & = & -2 |\alpha| |f(\theta)| 
z\big(t - \rho(\theta)/\omega_\mathrm{N}\big)  \quad 
\end{eqnarray}
where $f(\theta) = \chi_\C[\omega_\mathrm{N}] \chi^\ast_\C[-\omega_\mathrm{N}] ((\kappa/2 - \im \omega_\mathrm{N} ) \sin \tilde{\theta} - \Delta \cos \tilde{\theta} )$ and $\tilde{\theta} = \theta - \phi$ is the quadrature phase measured relative to the phase of the intensity quadrature. We observe that there is a phase shift $\rho(\theta)$ between the oscillator motion and the cavity field fluctuations, defined by $\mathrm{exp}(\im \rho(\theta)) = f(\theta)/|f(\theta)|$. Note again that in the bad cavity limit $\kappa \gg \omega_\mathrm{N}$, the choice $\theta = 2 \phi$ makes the prefactor ($|f(\theta)|$) vanish. 

In the general case, when examining the cross-correlation $\langle \delta X_{\theta_\RR}(t)  \delta Y_{\theta_\LL}(0) \rangle$, one finds that 
the term symmetric in time, which corresponds to the real part of $S[\omega]$, vanishes when $\rho(\theta_\RR) - \rho(\theta_\LL) = \pm \pi/2$, whereas the antisymmetric part, corresponding to the imaginary part of $S[\omega]$, vanishes when $\rho(\theta_\RR) - \rho(\theta_\LL) = 0,\pi$. The latter is always the case when $\theta_\RR = \theta_\LL$. The former criterion demands that the two quadrature phases satisfy
\begin{eqnarray}
  \label{eq:criterionThermalVanish}
  & & \left[ \left(\frac{\kappa}{2}\right)^2 + \omega_\mathrm{N}^2 + \Delta^2 \right] \cos(\tilde{\theta}_\RR - \tilde{\theta}_\LL)  -  \kappa \Delta \sin(\tilde{\theta}_\RR + \tilde{\theta}_\LL) \notag \\ & & - \left[\left(\frac{\kappa}{2}\right)^2 + \omega_\mathrm{N}^2 - \Delta^2 \right] \cos(\tilde{\theta}_\RR + \tilde{\theta}_\LL) = 0 \ . 
\end{eqnarray}
By choosing one of the quadratures, this equation gives the other quadrature for which $\mathrm{Re} \, S[\omega]$ will vanish. The physical interpretation is that the two quadratures $\delta X_{\theta_\RR}(t)$ and $\delta Y_{\theta_\RR}(t) $ then measure orthogonal quadratures of the mechanical oscillator's motion. 

In the toy example above, we made the assumption $z(t) \sim \cos(\omega_\mathrm{N} t)$. In reality, the oscillator motion is a noisy signal and not restricted to one frequency. If we still want the real part of the cross-correlation $S[\omega]$ to vanish, Equation (\ref{eq:criterionThermalVanish}) must be fulfilled for all frequencies $\omega_\mathrm{N}$. This is only possible if either $\theta_\RR$ or $\theta_\LL$ is equal to $\phi$, i.e., the intensity quadrature in the cavity. The natural choice is then $\theta_\RR = \phi$, since this can be achieved simply by replacing the homodyne detection on the right hand side with a photomultiplier and recording the intensity fluctuations. For $\Delta \neq 0$, the remainder of Equation (\ref{eq:criterionThermalVanish}) then dictates the other quadrature to be $\theta_\LL = 2 \phi$. To locate the correct $\theta_\LL$, one possibility is to drive the oscillator mechanically and look for the quadrature phase where the strong signal from the driven oscillator disappears in $\mathrm{Re} \, S[\omega]$. This null can be maintained via a servo loop throughout the course of the experiment provided that the external mechanical drive is at a frequency well away from the frequencies of interest.

Let us also mention that for a mechanical oscillator with a high quality factor $\omega_\M/\gamma$, the requirement that one of the quadrature phases must be $\phi$ can possibly be relaxed by only demanding Equation (\ref{eq:criterionThermalVanish}) to be valid for $\omega_\mathrm{N} = \omega_\M$. In that case, the thermal contribution to $\mathrm{Re} \, S[\omega]$ only approximately vanishes for frequencies close to the mechanical frequency. The calibration procedure mentioned above should also work in that case, but then only for drive frequencies close to the mechanical resonance frequency. Using $\theta_\RR = \phi$ and $\theta_\LL = 2 \phi$ does however also have the advantage that both angles are known, which is useful when trying to fit theoretical expressions to experimental results.

\subsection{General expressions}
\label{sec:GenExpressions}

\begin{figure}[htbp]
\begin{center} \includegraphics[width=.46\textwidth]{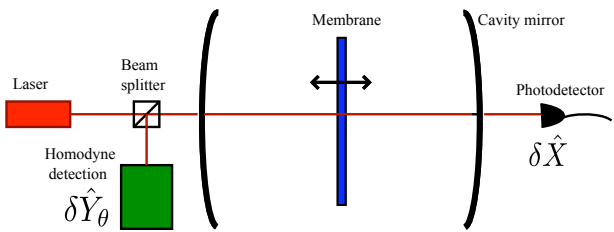}
  \caption{(color online). Modified version of the experiment considered, where the transmitted intensity fluctuations are measured by a photodetector.}
\label{fig:detection2}
\end{center}
\end{figure}

From the above discussion of the thermal noise contribution, we can conclude that it is favorable to let $\delta \Xhat_{\theta_\RR}$ be the fluctuations in the intensity quadrature, both for zero and nonzero detuning $\Delta$. This means that $\theta_\RR = \phi$, and we will restrict ourselves to this situation from now on. We also rename the quadratures by letting $\delta \Xhat_{\phi} \rightarrow \delta \Xhat$ and $\delta \Yhat_{\theta_\LL} \rightarrow \delta \Yhat_{\theta}$. Hence, we consider a modified version of the experiment, shown in Figure \ref{fig:detection2}, where the intensity fluctuations $\delta \Xhat$ are measured by a photodetector.
Let us however mention that the general expressions for the contributions to $S[\omega]$ presented below are valid also for arbitrary $\theta_\RR$, but with different coefficients.

In terms of $\zhat$, $\xihat_i$, $\delta x$ and $\delta y$, the correlation function $\langle \delta \Xhat[\omega] \delta \Yhat_\theta[\omega'] \rangle$ can be written as a sum of five contributions:
\begin{eqnarray}
  \label{eq:dXdYcont}
  & & \langle \delta \Xhat[\omega] \delta \Yhat_\theta[\omega'] \rangle  =  \langle \delta \Xhat[\omega] \delta \Yhat_\theta[\omega'] \rangle_\mathrm{q,q} \\
 &  & \qquad + \langle \delta \Xhat[\omega] \delta \Yhat_\theta[\omega'] \rangle_\mathrm{cl,cl} 
 +  \langle \delta \Xhat[\omega] \delta \Yhat_\theta[\omega'] \rangle_\mathrm{q,z} \nonumber \\
& & \qquad + \langle \delta \Xhat[\omega] \delta \Yhat_\theta[\omega'] \rangle_\mathrm{cl,z} + \langle \delta \Xhat[\omega] \delta \Yhat_\theta[\omega'] \rangle_\mathrm{z,z} \ . \nonumber 
\end{eqnarray}
Here, the label ``q'' refers to the quantum fields $\xihat_i$, the label ``cl'' refers to the classical fields $\delta x, \delta y$, and ``z'' refers to the position $\zhat$ of the mechanical oscillator. We have grouped the terms in this way to identify the signature of RPSN. The first term contains correlators of the kind $\langle \xihat_i[\omega] \xihat_i^\dagger[\omega'] \rangle$, with $i = \mathrm{L,R,M}$. It turns out that $\langle \delta \Xhat[\omega] \delta \Yhat_\theta[\omega'] \rangle_\mathrm{q,q} = 0$. Thus, in the absence of the nonlinearity introduced by the radiation-pressure-induced motion of the mechanical oscillator, there is no correlation between quantum noise in the outgoing quadratures on each side. The second term is the correlation between the classical laser noise in the two quadratures, containing the correlators $\langle \delta x[\omega] \delta x[\omega'] \rangle$ and $\langle \delta y[\omega] \delta y[\omega'] \rangle$. The third term, $\langle \delta \Xhat[\omega] \delta \Yhat_\theta[\omega'] \rangle_\mathrm{q,z}$, contains correlations between the quantum vacuum noise and the oscillator position, $\langle \xihat_i[\omega] z[\omega'] \rangle$, {\it which are the optomechanical correlations that we would like to detect}. The term $\langle \delta \Xhat[\omega] \delta \Yhat_\theta[\omega'] \rangle_{\mathrm{cl,z}}$ contains correlation functions of the kind $\langle \delta x[\omega] \zhat[\omega'] \rangle$, which are also optomechanical correlations, but due to classical noise in the drive, not photon shot noise. The last term ($\sim \langle \zhat[\omega] \zhat[\omega'] \rangle$), discussed in the previous section, is proportional to the position spectral density and appears as a result of detecting the mechanical oscillator position fluctuations in both quadratures. Although it vanishes at zero detuning $\Delta$, this term is generically non-zero.

By using Equations (\ref{eq:etaProperties}), (\ref{eq:xiProperties}), (\ref{eq:classProperties}) and (\ref{eq:zSolution}), we can evaluate the expectation values and write the cross-correlation $S[\omega]$ as
\begin{equation}
  \label{eq:Sparts}
  S[\omega] = S_\mathrm{q,z}[\omega] + S_\mathrm{cl,cl}[\omega] + S_\mathrm{cl,z}[\omega] + S_\mathrm{z,z}[\omega] \ ,
\end{equation}
where the terms correspond to the terms in Equation (\ref{eq:dXdYcont}). The correlation function $S[\omega]$ is a complex quantity, where the real(imaginary) part is symmetric(antisymmetric) in $\omega$. We will later use the abbreviations
\begin{eqnarray}
  \label{eq:ReImdef}
  R[\omega] & = & \mathrm{Re} \, S[\omega] \\
  I[\omega] & = & \mathrm{Im} \, S[\omega] \nonumber
\end{eqnarray}
and similarly for the various contributions in Equation (\ref{eq:Sparts}), i.e., $R_\mathrm{q,z}[\omega] = \mathrm{Re} \, S_\mathrm{q,z}[\omega]$ etc. 

The general expression for $S[\omega]$ can be found in Appendix \ref{app:genExp}. We are only interested in its behaviour around the mechanical frequency $\omega_\M$, where the oscillator is most susceptible to the radiation pressure noise. Considering a mechanical oscillator with a high quality factor, we assume $\tilde{\gamma} \ll \kappa, \omega_\M$ and restrict ourselves to frequencies where $|\omega - \omega_\M| \ll \kappa, \omega_\M$. This greatly simplifies the expressions. The (z,z)-contribution becomes
\begin{equation}
  \label{eq:SzzSimplified}
  S_\mathrm{z,z}[\omega] = \frac{R^\mathrm{(th)}_1 + \im I^\mathrm{(th)}_1}{\left(\omega - \tilde{\omega}_\M\right)^2 + \left(\tilde{\gamma}/2\right)^2}  \ .
\end{equation}
The real constants $R^\mathrm{(th)}_1$ and $I^\mathrm{(th)}_1$ are given in Equation (\ref{eq:Kthdef}) of Appendix \ref{app:genExp}. This means that in the vicinity of the mechanical frequency $\omega_\M$, both the real and imaginary parts of $S_\mathrm{z,z}[\omega]$ are Lorentzians centered around $\tilde{\omega}_\M$ with width $\tilde{\gamma}$. This is as expected, since $S_\mathrm{z,z}[\omega]$ originates from the spectral density of the mechanical oscillator. This term contains the thermal noise contribution to the cross-correlator $S[\omega]$, as the spectral density of the oscillator is typically dominated by thermal fluctuations. 

The term we would like to detect is
\begin{eqnarray}
  \label{eq:SqzSimplified}
  S_\mathrm{q,z}[\omega] = \frac{R^\mathrm{(q)}_1 + \im I^\mathrm{(q)}_1 + R^\mathrm{(q)}_2 \left(\omega - \tilde{\omega}_\M\right)}{\left(\omega - \tilde{\omega}_\M\right)^2 + \left(\tilde{\gamma}/2\right)^2}  \ ,
\end{eqnarray}
where the real constants $R^\mathrm{(q)}_1,R^\mathrm{(q)}_2,I^\mathrm{(q)}_1$ can be found in Equation (\ref{eq:Kqdef}). The imaginary part of $S_\mathrm{q,z}[\omega]$ is also a Lorentzian. The real part is a sum of two terms, one proportional to the above-mentioned Lorentzian and one term which changes sign at $\tilde{\omega}_\M$. We note at this point the important fact that when $|R^\mathrm{(q)}_1/R^\mathrm{(q)}_2|$ is small compared to $\tilde{\gamma}$, the real parts of $S_\mathrm{q,z}[\omega]$ and $S_\mathrm{z,z}[\omega]$ are qualitatively different and in principle distinguishable. The term containing correlations between classical noise in the drive and oscillator position becomes
\begin{eqnarray}
  \label{eq:SclzSimplified}
  S_\mathrm{cl,z}[\omega] = \frac{R^\mathrm{(cl)}_1 + \im I^\mathrm{(cl)}_1 + (R^\mathrm{(cl)}_2 + \im I^\mathrm{(cl)}_2) \left(\omega - \tilde{\omega}_\M\right)}{\left(\omega - \tilde{\omega}_\M\right)^2 + \left(\tilde{\gamma}/2\right)^2}  .
\end{eqnarray}
It is similar to $S_\mathrm{q,z}[\omega]$, but with different coefficients and an additional term in the imaginary part. An important reason for the difference between the quantum and classical contributions $S_\mathrm{q,z}[\omega]$ and $S_\mathrm{cl,z}[\omega]$ is that classical noise enters the cavity only from the left hand side, whereas quantum noise enters through all ports. Thus, an asymmetric cavity with $\kappa_\RR > \kappa_\LL$ can be favorable in terms of increasing the relative importance of the quantum versus classical contributions. Apart from this, there is no strong dependence on the relative size of $\kappa_\LL$, $\kappa_\RR$ and $\kappa_\M$. All plots in this article therefore refer to the case $\kappa_\M = 0$ and $\kappa_\LL = \kappa_\RR = \kappa/2$. The term $S_\mathrm{cl,cl}[\omega]$ has no sharp features around the mechanical frequency and produces only a smooth background. 

The detection of RPSN now comes down to being able to identify the presence of the term $S_\mathrm{q,z}[\omega]$ in the total signal $S[\omega]$ being measured.

\section{Quantum versus thermal contribution}
\label{sec:quanttherm}

\begin{figure}[htbp]
\begin{center}\includegraphics[width=.46\textwidth]{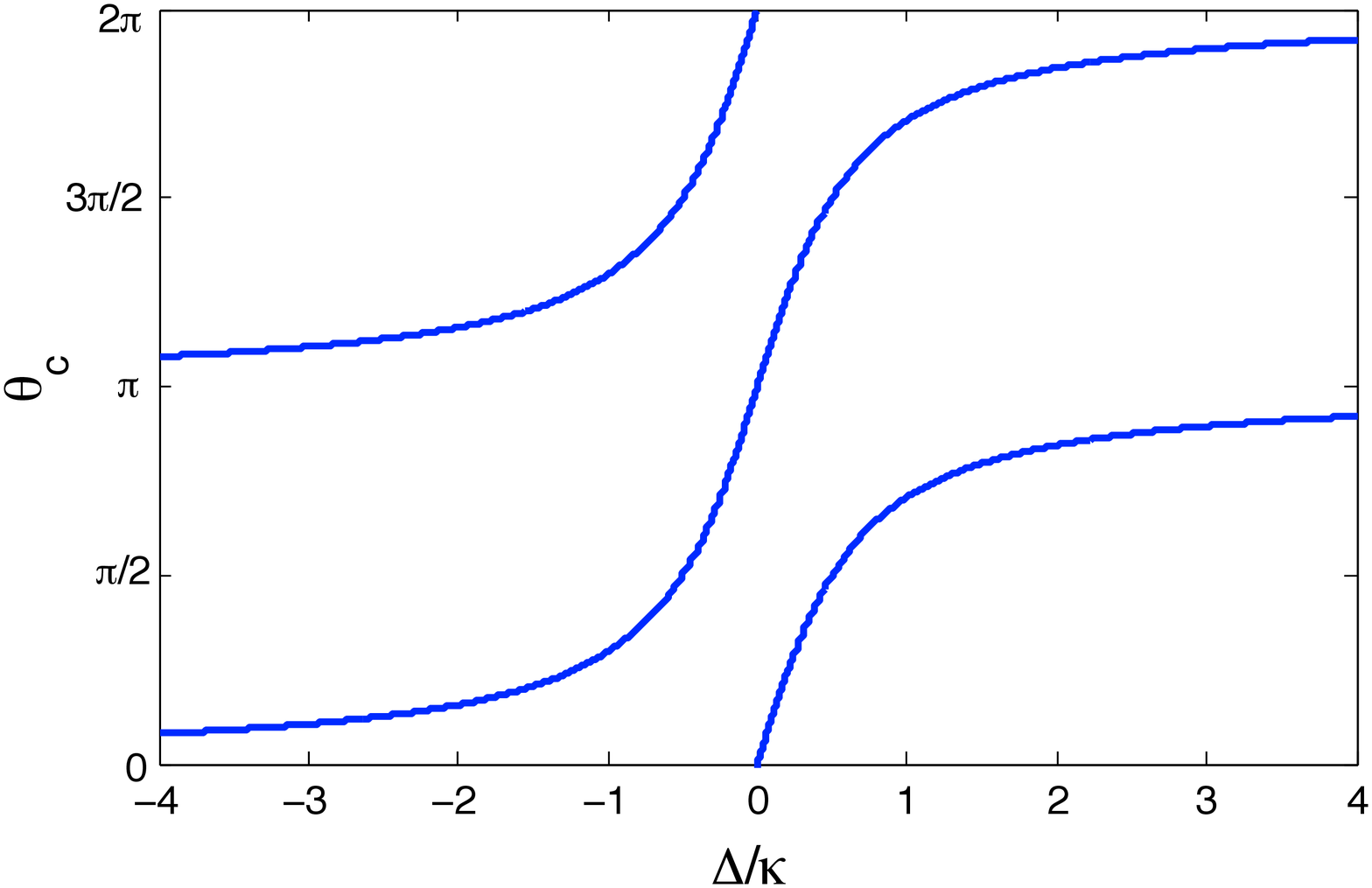}
\caption{(color online). The critical angle $\theta_c$ as a function of detuning $\Delta$. The angle $\theta_c$ is multivalued, since $\theta \rightarrow \theta + \pi$ gives $\delta \Yhat_\theta(t) \rightarrow - \delta \Yhat_\theta(t)$ and hence $S[\omega] \rightarrow -S[\omega]$.}
\label{fig:thetaCrit}
\end{center}
\end{figure}

In this section, we ignore classical noise in the drive, such that $S_\mathrm{cl,z}[\omega] = S_\mathrm{cl,cl}[\omega] =  0$, and focus on the two remaining contributions. We have observed that $R_\mathrm{z,z}[\omega]$ and $R_\mathrm{q,z}[\omega]$ in principle can be distuingished even at $\Delta \neq 0$, whereas the imaginary parts $I_\mathrm{z,z}[\omega]$ and $I_\mathrm{q,z}[\omega]$ are indistinguishable. However, $R_\mathrm{z,z}[\omega]$ is proportional to the mean phonon number $n_\M$, which, depending on the temperature $T$ of the mechanical bath, can be a macroscopic number. The contribution $R_\mathrm{q,z}[\omega]$ will therefore be negligible in most cases. In other words, the fluctuations of the oscillator due to the mechanical bath are usually much larger than those caused by a fluctuating photon number in the cavity, and consequently the latter is negligible. However, we are not measuring the position fluctuations directly, but rather the cross-correlations between the outgoing quadratures $\delta \Xhat$ and $\delta \Yhat_\theta$. In Section \ref{sec:Thermal}, we observed that at special points in parameter space the term $R_\mathrm{z,z}[\omega]$ vanishes. Heidmann {\it et al.}~\cite{Heidmann1997ApplPhysB} noted that this occurs at resonance, i.e., $\Delta = 0$, where $\delta \Xhat$ is independent of $\zhat$. We note that for non-zero detuning, it also occurs at the critical angle $\theta = \theta_c$, where $\theta_c = 2 \phi$ modulo $\pi$ and given by 
\begin{equation}
  \label{eq:thetaCritdef}
  \theta_c = \arctan \frac{\Delta/\kappa}{1/4 - \left(\Delta/\kappa\right)^2} + k \pi \ , \ k \in \mathbb{Z} \ ,
\end{equation}
This equation can be viewed in two ways. It gives the appropriate detuning $\Delta$ to make $R_\mathrm{z,z}[\omega]$ vanish for a given cavity linewidth $\kappa$ and angle $\theta$. For example, when $\theta = \pi/2$, this detuning is $|\Delta| = \kappa/2$. Equivalently, Equation (\ref{eq:thetaCritdef}) tells us which quadrature to measure on the left hand side of the cavity for a given $\Delta/\kappa$. Figure \ref{fig:thetaCrit} shows the critical angle $\theta_c$ as a function of detuning $\Delta$. 

Although the criteria for vanishing $R_\mathrm{z,z}[\omega]$ can be determined quite easily, it does not mean that experimental detection of $R_\mathrm{q,z}[\omega]$ becomes straightforward. There is always an uncertainty in experimental parameters as well as drift and fluctuations in laser frequency. The question is whether one can get sufficiently close to these criteria in order to claim that the correlation $R_\mathrm{q,z}[\omega]$ has been detected. In the following, we therefore aim to determine what sufficiently close means in quantitative terms.

\subsection{Zero detuning}

The unwanted contribution $R_\mathrm{z,z}[\omega]$ changes sign with $\Delta$ and vanishes when $\Delta = 0$. One might therefore be able to make $R_\mathrm{z,z}[\omega]$ small enough that $R_\mathrm{q,z}[\omega]$ is the dominant contribution to the cross-correlation measurement by choosing the detuning as close to $\Delta = 0$ as possible. Of course, a fluctuating and/or drifting $\Delta$ might pose a challenge in an actual experiment.

In the limit $|\Delta| \ll \kappa, \omega_\M$, the ratio $|R^\mathrm{(q)}_1/(R^\mathrm{(q)}_2 \tilde{\gamma})| \rightarrow 0$, such that the quantum contribution $R_\mathrm{q,z}[\omega]$ will have a sign change at $\tilde{\omega}_\M$. In fact, $S_\mathrm{q,z}$ takes exactly the form of the simple force-position correlation discussed in Section \ref{sec:Motivation}, but with $\omega_\M \rightarrow \tilde{\omega}_\M$ and $\gamma \rightarrow \tilde{\gamma}$. Also, $S_\mathrm{q,z}[\omega]$ is proportional to $\sin \theta$ in this limit, such that $\theta = \pi/2$ will maximize the signal, as was also noted in Ref. \cite{Heidmann1997ApplPhysB}. 

\begin{figure}[htbp]
\begin{center} \includegraphics[width=.46\textwidth]{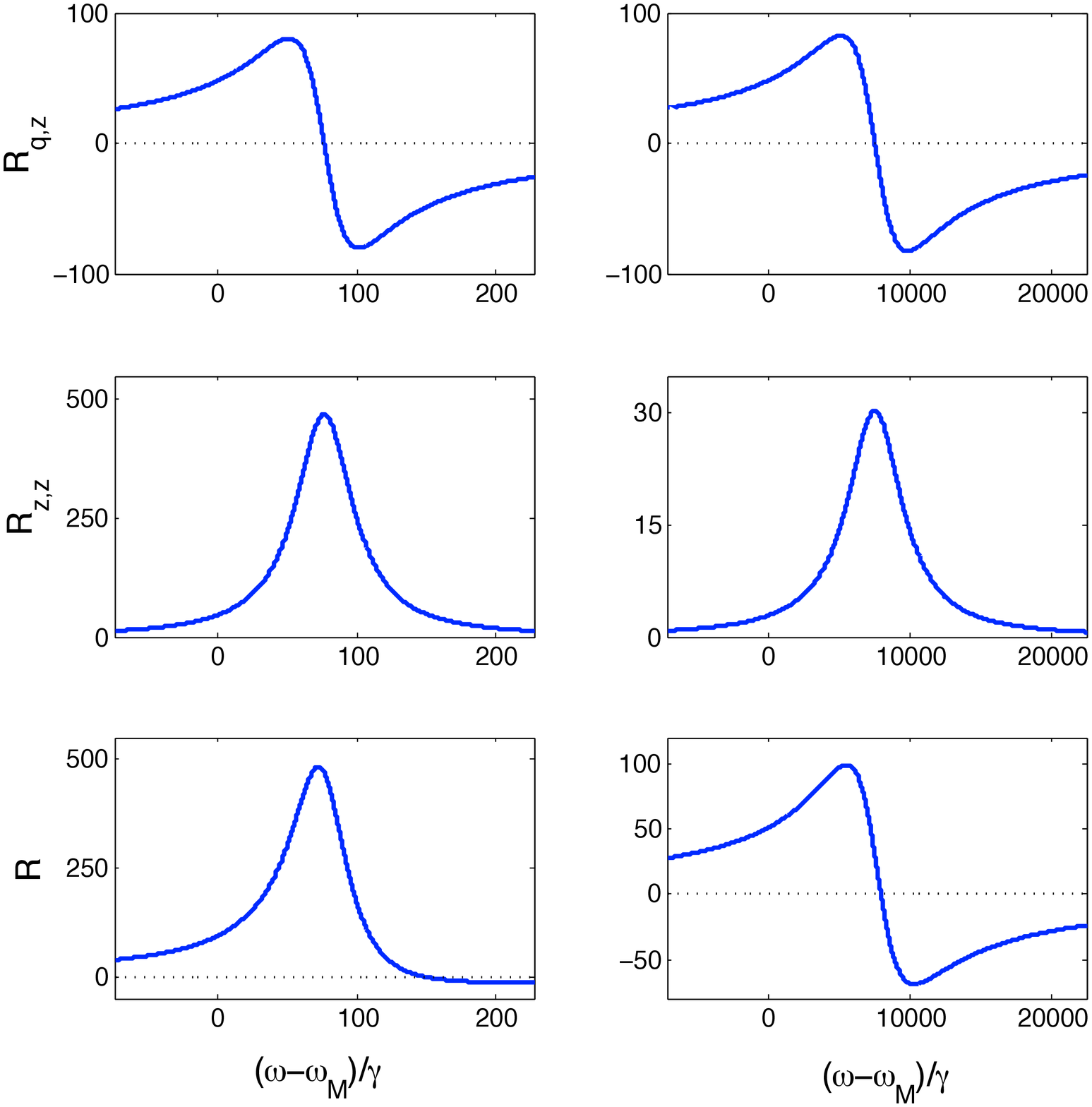}
\caption{(color online). The contributions $R_\mathrm{q,z}[\omega]$ {\it (upper panel)} and $R_\mathrm{z,z}[\omega]$ {\it (middle panel)}, and their sum $R[\omega]$ {\it (lower panel)}. We have used $\Delta = -0.01 \kappa$, $\omega_\M/\kappa = 3.2$, $\omega_M/\gamma = 10^6$, $A/\omega_\M = 1.15 \cdot 10^{-6}$ and $\theta = \pi/2$. The temperature is 4.2 K. The mean number of photons in the cavity ($n_\mathrm{photon} = |\bar{a}|^2$) is $10^{10}$ {\it (left panel)} and $10^{12}$ {\it (right panel)}. In the first case, the total signal is dominated by $R_\mathrm{z,z}[\omega]$, whereas in the second case $R_\mathrm{q,z}[\omega]$ dominates. In the absence of classical laser noise, a measured signal as in the lower right panel is a signature of RPSN.}
\label{fig:RDelta0}
\end{center}
\end{figure}

 Figure \ref{fig:RDelta0} shows the two contributions $R_\mathrm{q,z}[\omega]$ and $R_\mathrm{z,z}[\omega]$, as well as the total signal $R[\omega] = R_\mathrm{q,z}[\omega] + R_\mathrm{z,z}[\omega]$, for a detuning $\Delta = -0.01\kappa$. The parameters chosen are relevant to the present membrane-in-the-middle setup \cite{Thompson2008Nature,Jayich2008NJP}. In the right panel, the mean number of photons $n_\mathrm{photon} = |\bar{a}|^2$ in the cavity is 100 times larger than in the left panel, corresponding to a difference in laser power. The abscissa shows the frequency deviation $\omega - \omega_\M$ in units of the bare mechanical damping $\gamma$. In the case displayed in Figure \ref{fig:RDelta0}, we are in the regime where $\gamma_\mathrm{opt},\delta \omega_\M \gg \gamma$, where $\delta \omega_\M$ is the ``optical spring'' frequency shift and the linewidth increase $\gamma_\mathrm{opt}$ is a result of optical cooling of the mechanical motion associated with the detuning being negative \cite{Marquardt2007PRL}. In the top panel, $R_\mathrm{q,z}[\omega]$ is shown, and we observe that it is dominated by the $R_2^\mathrm{(q)}$-term in Equation (\ref{eq:SqzSimplified}). In the middle panel, the Lorentzian $R_\mathrm{z,z}[\omega]$ is shown. Increasing the photon number by a factor of 100, i.e., going from the left to the right panel, we see that both $R_\mathrm{q,z}[\omega]$ and $R_\mathrm{z,z}[\omega]$ become wider and are shifted further from $\omega_\M$ due to the increase in $\gamma_\mathrm{opt}$ and $\delta \omega_\M$. We observe that the peak-to-peak value of $R_\mathrm{q,z}[\omega]$ is largely unchanged, whereas the height of $R_\mathrm{z,z}[\omega]$ is smaller when the photon number is increased. The latter is due to increased optical cooling of the mechanical oscillator. In the lower panel, the total signal $R[\omega]$ is shown. On the left hand side, the contribution $R_\mathrm{z,z}[\omega]$ is dominant and the total signal looks almost like a Lorentzian, albeit with some asymmetry. On the right hand side, $R_\mathrm{q,z}[\omega]$ dominates, and the total signal has a sign change and is clearly not a Lorentzian. Assuming there is no classical noise in the drive, a signal $R[\omega]$ as in the lower right panel of Figure \ref{fig:RDelta0} is a signature of correlations between photon shot noise in the cavity and position fluctuations of the mechanical oscillator.

From Equations (\ref{eq:SqzSimplified}) and (\ref{eq:SzzSimplified}), we can derive the peak-to-peak value $P_\mathrm{q}$ of the contribution $R_\mathrm{q,z}[\omega]$ and the height $M$ of the Lorentzian $R_\mathrm{z,z}[\omega]$. To be able to detect correlations between photon shot noise and oscillator position, an important figure of merit is the ratio $P_\mathrm{q}/M$. We find that for $|\Delta| \ll \kappa, \omega_\M$,
\begin{equation}
  \label{eq:PqoverMD0}
  \frac{P_\mathrm{q}}{M} = \frac{\left(\kappa/2\right)^2 + \omega_\M^2}{2 \left(n_\M + 1/2\right) \kappa |\Delta|} \ ,
\end{equation}
where the effective phonon number $n_\M$ is given in Equation (\ref{eq:nMeff}). When this ratio becomes larger than one, the total signal $R[\omega]$ is dominated by $R_\mathrm{q,z}[\omega]$. In Figure \ref{fig:RDelta0}, this ratio is 0.34 (left panel) and 5.42 (right panel).

Let us assume that the parameters $\omega_\M, \kappa, \gamma, n_\mathrm{th}$ are fixed and we are free to change the detuning $\Delta$ and the optomechanical coupling $|\alpha|$. The latter can be varied by changing the input power and hence the intracavity photon number. First, consider the regime $\gamma_\mathrm{opt} \ll \gamma$, where $n_\M \approx n_\mathrm{th}$. In that case $P_\mathrm{q}/M$ is simply inversely proportional to $|\Delta|$. In the opposite regime, $\gamma_\mathrm{opt} \gg \gamma$, Equation (\ref{eq:PqoverMD0}) becomes
\begin{equation}
  \label{eq:PqoverMD0approx}
  \frac{P_\mathrm{q}}{M} \approx \frac{2\omega_\M |\alpha|^2}{\gamma n_\mathrm{th} \left[\left(\kappa/2\right)^2 + \omega_\M^2\right] +\kappa |\alpha|^2} \ ,
\end{equation}
when assuming $n_\M \gg 1$. Note that in this regime ($|\Delta| \ll \kappa, \omega_\M$, $\gamma \ll \gamma_\mathrm{opt}$), the figure of merit is independent of $\Delta$. The reason is that an increase of $\Delta$ is compensated by additional cooling of the oscillator. The ratio can be made larger by increasing $|\alpha|$ \footnote{Note however that $|\alpha|$ is limited by our weak-coupling assumption $\tilde{\gamma} \ll \kappa, \omega_\M$.}, i.e., input power, and has a maximal value of $2\omega_\M/\kappa$. It is worth noting that a decrease of $|\Delta|$ in this regime does not necessarily pay off in an increased $P_\mathrm{q}/M$. For example, a decrease of $|\Delta|$ by a factor of 10 does not significantly change this ratio in the examples shown in Figure \ref{fig:RDelta0}. However, the overall signal gets larger and one might become less sensitive to technical noise. In addition, when $|\Delta|$ gets small enough, we eventually reach the point where $\gamma_\mathrm{opt}$ and $\gamma$ become comparable and Equation (\ref{eq:PqoverMD0approx}) loses validity.
 
\subsection{Finite detuning}

At finite detuning $\Delta$, the signal $R_\mathrm{z,z}[\omega]$ vanishes when $\theta = \theta_c$. We now investigate whether $R_\mathrm{q,z}[\omega]$ is sufficiently distinguishable from $R_\mathrm{z,z}[\omega]$ also in this case, and how small $|\theta - \theta_c|$ must be for $R_\mathrm{q,z}[\omega]$ to dominate.

We noted above that the correlation we seek to detect, $R_\mathrm{q,z}[\omega]$, is distinguishable from $R_\mathrm{z,z}[\omega]$ when $|R^\mathrm{(q)}_1/(R^\mathrm{(q)}_2 \tilde{\gamma})| \ll 1$. For $\theta \approx \theta_c$, this requirement becomes
\begin{equation}
  \label{eq:Kq2Kq1ratio}
\frac{\omega_\M \kappa}{|\left(\kappa/2\right)^2+\Delta^2-\omega_\M^2|} \ll 1 \ .
\end{equation}
This can be satisfied in several ways, for example in the resolved sideband limit $\omega_\M \gg \kappa$ with $|\Delta| \ll \omega_\M$.  In an experiment where the oscillator motion is dominated by thermal fluctuations, the quadrature $\theta = \theta_c$ can be located by looking for an overall sign change in the total signal $R[\omega]$, either by varying $\theta$ or, if $\theta$ is fixed, by varying $\Delta$. Exactly at $\theta = \theta_c$, one should be able to see the asymmetric features originating from $R_\mathrm{q,z}[\omega]$. As mentioned earlier, the calibration process for locating $\theta_c$ can be improved by strongly driving the oscillator mechanically to temporarily amplify the (z,z)-contribution.

\begin{figure}[htbp]
\begin{center} \includegraphics[width=.46\textwidth]{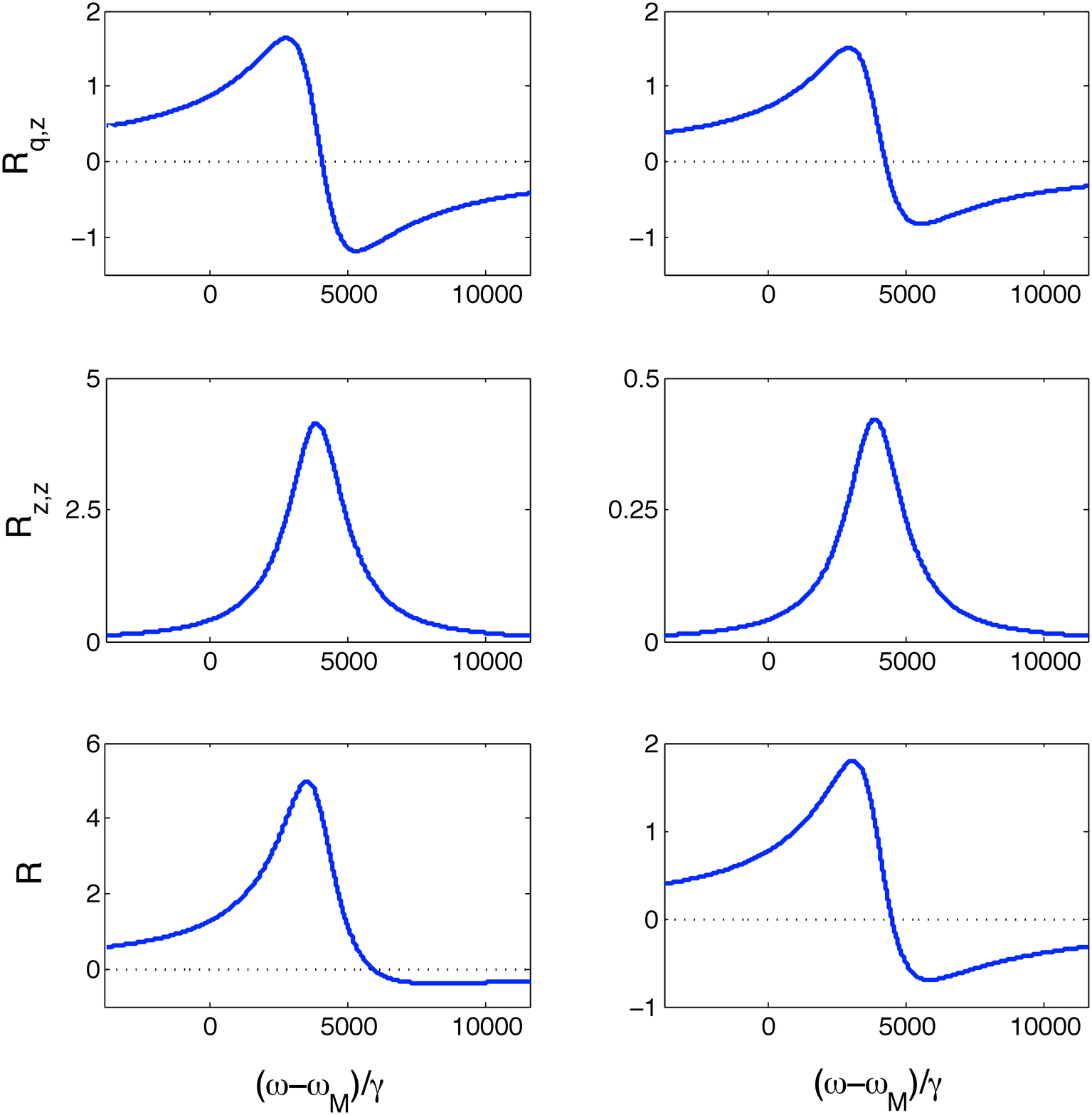}
\caption{(color online). The contributions $R_\mathrm{q,z}[\omega]$ {\it (upper panel)} and $R_\mathrm{z,z}[\omega]$ {\it (middle panel)}, and their sum $R[\omega]$ {\it (lower panel)}. We have used $\Delta = -0.5 \kappa$, $\omega_\M/\kappa = 3.2$, $\omega_M/\gamma = 10^6$ and $A/\omega_\M = 1.15 \cdot 10^{-6}$. The temperature is 4.2 K. The mean number of photons in the cavity ($n_\mathrm{photon} = |\bar{a}|^2$) is $10^{10}$. The angle $\theta$ is $\theta_c-\pi/10$ {\it (left panel)} and $\theta_c-\pi/100$ {\it (right panel)}. In the first case, the total signal is dominated by $R_\mathrm{z,z}[\omega]$, whereas in the second case, $R_\mathrm{q,z}[\omega]$ dominates.}
\label{fig:RDeltam05}
\end{center}
\end{figure}

In Figure \ref{fig:RDeltam05}, we show the real part of the cross-correlation $R[\omega]$ and its contributions for $\Delta = - \kappa/2$. In this case, $\theta_c = \pi/2$. In the left panel $\theta = \theta_c - \pi/10$, and in the right panel $\theta = \theta_c - \pi/100$. We see that the unwanted contribution $R_\mathrm{z,z}[\omega]$ dominates in the first case, whereas $R_\mathrm{q,z}[\omega]$ is dominant when $\theta$ gets closer to $\theta_c$.

Again, the important figure of merit is the ratio between the peak-to-peak value $P_\mathrm{q}$ of the contribution $R_\mathrm{q,z}[\omega]$ and the height $M$ of the Lorentzian $R_\mathrm{z,z}[\omega]$. For $\delta \theta = |\theta - \theta_c| \ll 1$, it is
\begin{equation}
  \label{eq:PqoverMTheta}
  \frac{P_\mathrm{q}}{M} = \frac{\sqrt{(\omega_\M \kappa)^2 + \left[(\kappa/2)^2+\Delta^2 - \omega_\M^2\right]^2}}{4(n_\M + 1/2) \left[(\kappa/2)^2+\Delta^2\right] \delta\theta} .
\end{equation}
In Figure \ref{fig:RDeltam05}, this ratio is 0.54 (left panel) and 5.44 (right panel) \footnote{The value for the left panel is inaccurate, since $\delta\theta = \pi/10$ is a bit large for Equation (\ref{eq:PqoverMTheta}) to be valid.}. 

\begin{figure}[htbp]
\begin{center} \includegraphics[width=.46\textwidth]{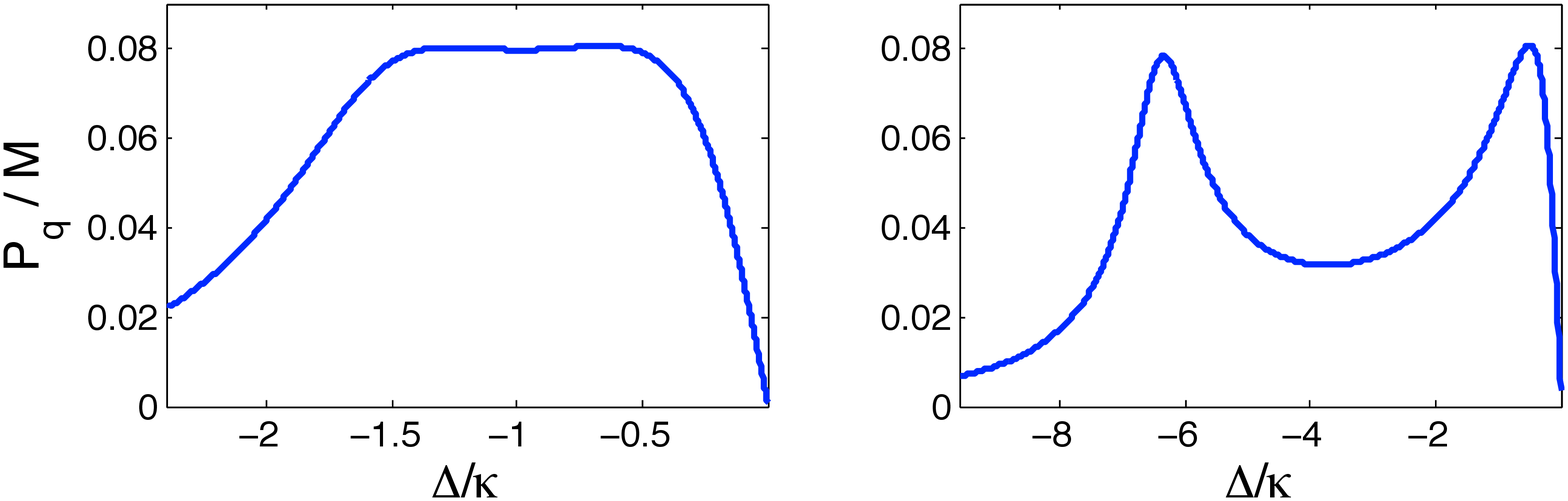}
\caption{(color online). The ratio between the peak-to-peak value $P_\mathrm{q}$ of $R_\mathrm{q,z}[\omega]$ and the height $M$ of $R_\mathrm{z,z}[\omega]$ for $\omega_M/\kappa = 1.6$ {\it (left panel)} and $\omega_\M/\kappa = 6.4$ {\it (right panel)}. The deviation from the critical angle $\theta_c$ is $\delta \theta = \pi/100$, $\omega_M/\gamma = 10^6$, $A/\omega_\M = 1.15 \cdot 10^{-6}$, $n_\mathrm{photon} = 10^{10}$, and the temperature is $300$ K.}
\label{fig:Ratio300K}
\end{center}
\end{figure}

\begin{figure}[htbp]
\begin{center} \includegraphics[width=.46\textwidth]{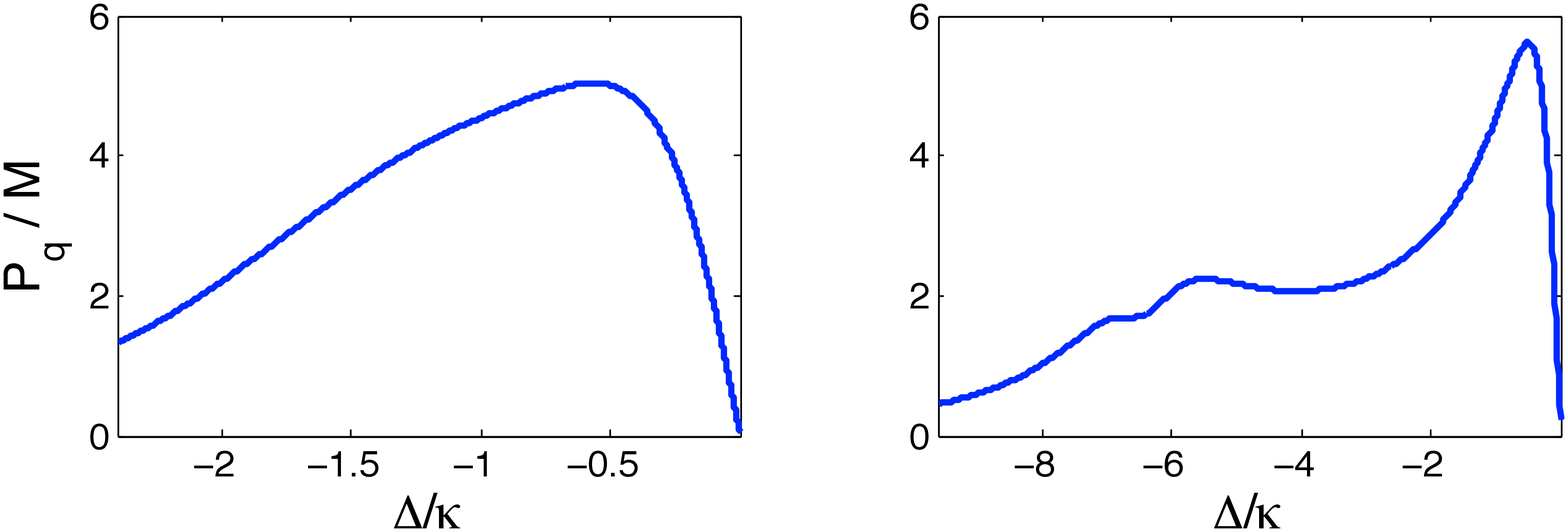}
\caption{(color online). The ratio between the peak-to-peak value $P_\mathrm{q}$ of $R_\mathrm{q,z}[\omega]$ and the height $M$ of $R_\mathrm{z,z}[\omega]$ for $\omega_M/\kappa = 1.6$ {\it (left panel)} and $\omega_\M/\kappa = 6.4$ {\it (right panel)}. The deviation from the critical angle $\theta_c$ is $\delta \theta = \pi/100$, $\omega_M/\gamma = 10^6$, $A/\omega_\M = 1.15 \cdot 10^{-6}$, $n_\mathrm{photon} = 10^{10}$, and the temperature is $4.2$ K.}
\label{fig:Ratio4k2K}
\end{center}
\end{figure}

Figure \ref{fig:Ratio300K} shows the ratio (\ref{eq:PqoverMTheta}) as a function of $\Delta$ at a temperature of 300 K with $\delta \theta = \pi/100$. We have used $\omega_\M/\kappa = 1.6$ and $\omega_\M/\kappa = 6.4$ in the left and right panel, respectively. Figure \ref{fig:Ratio4k2K} shows the same ratio, only at $T = 4.2$ K, giving significantly higher values. We observe that for small $\omega_\M/\kappa$, the ratio is not very dependent on $\Delta$, but for larger values of $\omega_\M/\kappa$ it has a peak at $\Delta = - \kappa/2$. For high temperatures, it also has a peak at $\Delta = -\omega_\M$. This is in accordance with the fact that cooling of the mechanical motion is most effective at $\Delta = -\omega_\M$ when $\omega_\M/\kappa$ is large \cite{Marquardt2007PRL,Wilson-Rae2007PRL}. Note however that the condition $\Delta = -\omega_\M$ with $\omega_\M \gg \kappa$ does not satisfy Equation (\ref{eq:Kq2Kq1ratio}).

From the above, we can conclude that it might be possible to observe the correlation between photon shot noise and oscillator position using a negative detuning $\Delta$ and $\theta = \theta_c$, especially in the resolved sideband regime where $\omega_\M/\kappa$ is large. We should point out that $\theta_c$ is susceptible to fluctuations in the detuning $\Delta$, as can be seen from Equation (\ref{eq:thetaCritdef}). From this point of view, it would be beneficial to choose $|\Delta|$ large compared to the cavity linewidth $\kappa$, since $\du \theta_c/\du \Delta \rightarrow 0$ when $|\Delta|/\kappa$ becomes large (see Figure \ref{fig:thetaCrit}). However, the choice of the detuning $\Delta$ is also limited by other requirements, such as Equation (\ref{eq:Kq2Kq1ratio}).

The potential insensitivity to laser frequency fluctuations for large $|\Delta|/\kappa$ could offer an advantage of this method versus performing the experiment at resonance. Another advantage is a lower effective phonon number $n_\M$ due to greater optical cooling of the mechanical motion. On the other hand, the reduced overall size of the signal, which reduces the signal to (technical) noise ratio, might be a disadvantage.

\section{Quantum versus classical contribution}
\label{sec:quantclass}

Even if a signal as in the lower right panel of Figures \ref{fig:RDelta0} and \ref{fig:RDeltam05} is detected, one needs to make sure that it represents a correlation between the mechanical oscillator position and photon shot noise, not classical laser noise. We now ignore the contribution $S_\mathrm{z,z}[\omega]$ and look at whether or not we can distinguish the classical and quantum contributions. We again focus on the two cases where the thermal signal $R_\mathrm{z,z}[\omega]$ vanishes, $\Delta = 0$ and, for $\Delta \neq 0$, $\theta = \theta_c$.

\subsection{Zero detuning}

We saw earlier that $R_\mathrm{q,z}[\omega]$ has a sign change at $\tilde{\omega}_\M$ when $|\Delta| \ll \kappa,\omega_\M$, since $|R^\mathrm{(q)}_1/(R^\mathrm{(q)}_2 \tilde{\gamma})| \rightarrow 0$ in that limit. The corresponding ratio for the classical signal $R_\mathrm{cl,z}[\omega]$ becomes $|R^\mathrm{(cl)}_1/(R^\mathrm{(cl)}_2 \tilde{\gamma})| = \omega_\M/\kappa$ when $|\Delta| \ll \kappa,\omega_\M$. This means that in the good cavity limit $\omega_\M/\kappa \gg 1$ and close to resonance, $R_\mathrm{cl,z}[\omega]$ is approximately a Lorentzian (see Equation (\ref{eq:SclzSimplified})), and the classical and quantum contributions are distinguishable. This is a consequence of the cavity being two-sided. In the bad cavity limit $\omega_\M/\kappa \ll 1$, the classical and quantum signals $R_\mathrm{q,z}[\omega]$ and $R_\mathrm{cl,z}[\omega]$ cannot be distinguished, although it might still be possible to exclude the classical laser noise contribution from other types of measurement. 

The ratio between the peak-to-peak values of $R_\mathrm{q,z}[\omega]$ and $R_\mathrm{cl,z}[\omega]$ is
\begin{equation}
  \label{eq:PqPclDelta0}
  \frac{P_\mathrm{q}}{P_\mathrm{cl}} = \frac{\sqrt{(\kappa/2)^2+\omega_\M^2}}{4\kappa_\LL C_X}
\end{equation}
for $|\Delta| \ll \kappa,\omega_\M$. As motivated above, we observe that an asymmetric cavity with $\kappa_\RR > \kappa_\LL$ will reduce the relative contribution from classical noise in this case. Equation (\ref{eq:PqPclDelta0}) also suggests operating in the resolved sideband regime, $\omega_\M/\kappa \gg 1$. 

\begin{figure}[htbp]
\begin{center}
\includegraphics[width=.46\textwidth]{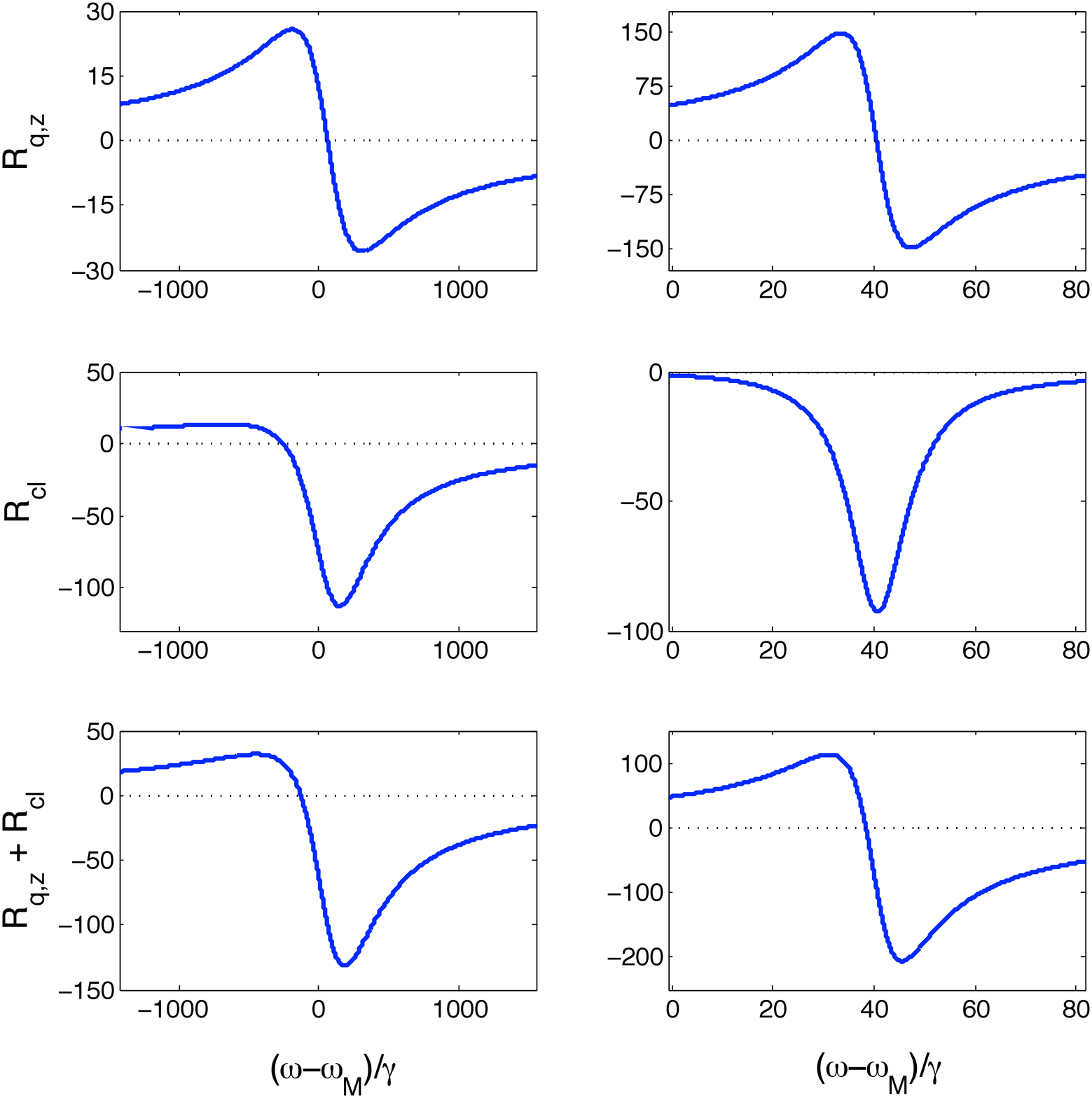}
\caption{(color online). The contributions $R_\mathrm{q,z}[\omega]$ {\it (upper panel)}, $R_\mathrm{cl}[\omega]$ {\it (middle panel)}, and their sum {\it (lower panel)}. The detuning is $\Delta = -0.01 \kappa$, $\omega_M/\gamma = 10^6$, $A/\omega_\M = 1.15 \cdot 10^{-6}$, $\theta = \pi/2$ and $n_\mathrm{photon} = 10^{10}$. The ratio between mechanical frequency and cavity linewidth is $\omega_\M/\kappa = 0.64$ {\it (left panel)} and $\omega_\M/\kappa = 6.4$ {\it (right panel)}. Note the qualitative difference between the two contributions in the latter case.}
\label{fig:RclDelta0F20000200000}
\end{center}
\end{figure}

The two points noted above are illustrated in Figure \ref{fig:RclDelta0F20000200000}, which shows the quantum contribution $R_\mathrm{q,z}[\omega]$, the classical contribution $R_\mathrm{cl}[\omega] = R_\mathrm{cl,z}[\omega] + R_\mathrm{cl,cl}[\omega]$, and their sum. We have used $\omega_\M/\kappa = 0.64$ and $\omega_\M/\kappa = 6.4$ in the left and right panels, respectively. The strength of the classical noise was chosen to be $C_X = C_Y = 1$, making it comparable to the quantum noise. In the left panel, we observe that the classical and quantum signals are of the same order, and it is hard to distinguish them. In the right panel, the classical contribution looks more like a (negative) Lorentzian, whose height is smaller than the peak-to-peak value of the quantum contribution. 

We should also mention that the imaginary parts $I_\mathrm{q,z}[\omega]$ and $I_\mathrm{cl,z}[\omega]$ are qualitatively different, since the former is simply a Lorentzian. Examining the imaginary part $I[\omega]$ could therefore provide a way of deciding whether the observed signal $R[\omega]$ is of quantum or classical nature. See Appendix \ref{sec:coeffResonance} for details.

\subsection{Finite detuning}

\begin{figure}[htbp]
\begin{center}
\includegraphics[width=.46\textwidth]{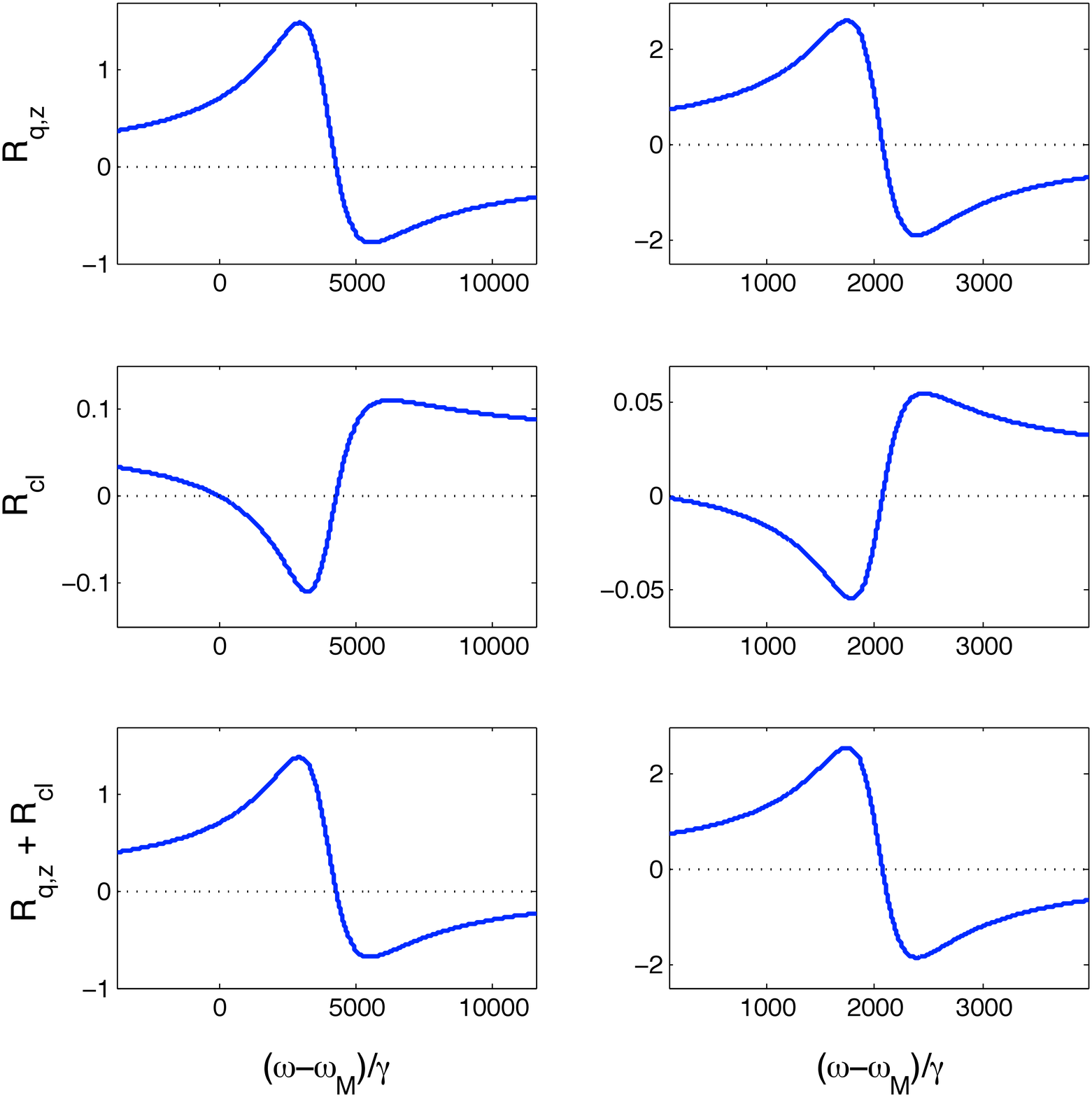}
\caption{(color online). The contributions $R_\mathrm{q,z}[\omega]$ {\it (upper panel)}, $R_\mathrm{cl}[\omega]$ {\it (middle panel)}, and their sum {\it (lower panel)}. The detuning is $\Delta = -\kappa/2$, $\omega_M/\gamma = 10^6$, $A/\omega_\M = 1.15 \cdot 10^{-6}$, $n_\mathrm{photon} = 10^{10}$ and $\theta = \theta_c$. The ratio between mechanical frequency and cavity linewidth is $\omega_\M/\kappa = 3.2$ {\it (left panel)} and $\omega_\M/\kappa = 6.4$ {\it (right panel)}.}
\label{fig:RclDeltam0k5F100000200000}
\end{center}
\end{figure}

For angles $\theta$ close to $\theta_c$, we know that the quantum signal $R_\mathrm{q,z}[\omega]$ is distinguishable from $R_\mathrm{z,z}[\omega]$ when Equation (\ref{eq:Kq2Kq1ratio}) is satisfied. In this case, the coefficients $R^\mathrm{(cl)}_1$ and $R^\mathrm{(cl)}_2$ are so complicated that an expression like Equation (\ref{eq:PqPclDelta0}) is not very helpful. However, the relative size of the quantum and classical contributions can always be examined by plotting the two contributions. Figure \ref{fig:RclDeltam0k5F100000200000} shows two examples, where $\Delta = -\kappa/2$, $\theta = \theta_c = \pi/2$, $C_X = C_Y = 1$, and where $\omega_\M/\kappa = 3.2$ (left panel) and $\omega_\M/\kappa = 6.4$ (right panel). 

We observe that the quantum contribution dominates in both cases, and that a large ratio $\omega_\M/\kappa$ seems to be favorable also here.

\section{Two optical modes}
\label{sec:twomodes}

As mentioned in the introduction, the most studied system in optomechanics is the one where an optical cavity mode is coupled to the motion of a movable end mirror \cite{Fabre1994PRA,Jacobs1994PRA,Metzger2004Nature,Gigan2006Nature,Arcizet2006Nature,Kleckner2006Nature,Hertzberg2010NatPhys,Groblacher2009Nature}. The membrane-in-the-middle geometry \cite{Thompson2008Nature,Jayich2008NJP} is an alternative setup where one avoids having to combine a high finesse cavity mirror and a delicate mechanical element. In both geometries, it is beneficial to make the intracavity photon number and hence the optomechanical coupling $|\alpha|$ as large as possible for the observation of RPSN. This can be achieved by increasing the power of the laser driving the cavity. However, one problem with this is that the mean radiation pressure on the mechanical oscillator increases, which leads to an increased shift in its equilibrium position. This eventually leads to a static instability \cite{Dorsel1983PRL,Jayich2008NJP} where more than one stable equilibrium position exist. This effect limits the amount of laser power that can be applied. 

It turns out that there is a way to perform the membrane-in-the-middle experiment with zero mean radiation pressure on the membrane. The trick is to drive two optical modes in the cavity whose coupling to the oscillator is of opposite sign. In other words, the two eigenmode frequencies would depend oppositely on the oscillator position \cite{Thompson2008Nature}. It should then be possible to increase the intracavity photon number without worrying about the static instability, and thus increasing the possibility of observing RPSN. Note that this is not possible in the movable mirror setup, where the coupling always has the same sign. If in addition, both modes are driven by the same laser, e.g., by utilizing an acousto-optic modulator, the classical noises in the two modes are correlated whereas the quantum noises are not. The classical noises in the two cavity modes will then try to force the mechanical oscillator in opposite directions in a synchronized way, resulting in a small net displacement. This can significantly reduce the oscillator fluctuations induced by classical laser noise, and thus improve the chances of observing the fluctuations induced by photon shot noise. We briefly discuss this setup below. Although it is somewhat specific to the membrane-in-the-middle geometry, it may be of interest in other setups as well \cite{Murch2008NatPhys,Eichenfield2009Nature}.

We now consider two optical modes $\ade$ and $\hat{b}$ both coupled to the mechanical oscillator,
\begin{eqnarray}
 \label{eq:hamiltonian2}
  \Hhat & = & \hbar \omega_\M \ccr \cde + \hbar \left( \omega_{\C,a} + A \zhat \right) \left(\acr \ade - \langle \acr \ade \rangle \right) \\
  & + & \hbar \left( \omega_{\C,b} - B \zhat \right) \left(\hat{b}^\dagger \hat{b} - \langle \hat{b}^\dagger \hat{b} \rangle \right)  + H_{\kappa_a} + H_{\kappa_b} + H_{\gamma}  \ , \notag
\end{eqnarray}
where the coupling constants $A$ and $B$ have the same sign. If $A\langle \acr \ade \rangle = B \langle \hat{b}^\dagger \hat{b} \rangle$, the mechanical oscillator's equilibrium position is the same as in the absence of driving the cavity modes. We again assume that the cavity is driven from the left, with frequencies $\omega_{\mathrm{D},a}$ and $\omega_{\mathrm{D},b}$ for the two modes $\hat{a}$ and $\hat{b}$, and define the detunings $\Delta_a = \omega_{\mathrm{D},a} - \omega_{\C,a}$ and $\Delta_b = \omega_{\mathrm{D},b} - \omega_{\C,b}$. We assume that the two cavity modes are well separated in frequency compared to their linewidths $\kappa_a$ and $\kappa_b$ and to the detunings $\Delta_{a}$ and $\Delta_b$, i.e., $|\omega_{\C,a}-\omega_{\C,b}| \gg \kappa_{a,b},|\Delta_{a,b}|$. 

There are now six input modes, $\hat{a}_{\mathrm{in},i}(t)$ and $\hat{b}_{\mathrm{in},i}(t)$, with $i = \LL,\, \RR,\, \M$. We write them as 
\begin{eqnarray}
  \label{eq:inputTwoModes}
  \hat{a}_{\mathrm{in},\LL}(t) & = & \ex^{-\im \omega_{\mathrm{D},a} t} \left( \bar{a}_\mathrm{in}(t) + \xihat_{a,\LL} \right) \\
  \hat{b}_{\mathrm{in},\LL}(t) & = & \ex^{-\im \omega_{\mathrm{D},b} t} \left( \bar{b}_\mathrm{in}(t) + \xihat_{b,\LL} \right) \nonumber \\
  \hat{a}_{\mathrm{in},i}(t) & = & \ex^{-\im \omega_{\mathrm{D},a} t} \xihat_{a,i}  \ , \ i = \RR,\M \nonumber \\
 \hat{b}_{\mathrm{in},i}(t) & = & \ex^{-\im \omega_{\mathrm{D},b} t} \xihat_{b,i} \ , \ i = \RR,\M \ , \nonumber
\end{eqnarray}
where the quantum noise operators $\xihat_{j,i}$ have the same properties as before.
If we assume that the classical noises in the two beams are correlated, the coherent state amplitudes are $\bar{a}_\mathrm{in}(t) = \bar{a}_0 + \delta x(t) + \im \delta y(t)$ and $\bar{b}_\mathrm{in}(t) = \bar{a}_\mathrm{in}(t) \, \bar{b}_0/\bar{a}_0$. We have assumed that $\bar{a}_0$ and $\bar{b}_0 $ are real. Also, since we are only interested in the noises $\delta x$ and $\delta y$ around the mechanical frequency, we can neglect any effects due to different path lengths for the two beams. We write the cavity amplitudes as a mean and a fluctuating part, $\hat{a} = \ex^{-\im \omega_{\mathrm{D},a} t} \left(\bar{a} + \dhat_a(t)\right)$ and $\hat{b} = \ex^{-\im \omega_{\mathrm{D},b} t} \left(\bar{b} + \dhat_b(t)\right)$. Linearization of the equations of motion gives
\begin{eqnarray}
  \label{eq:EOMTwoModes}
  \dot{\dhat}_a & = & - \left(\frac{\kappa_a}{2} - \im \Delta_a \right) \dhat_a - \im \alpha \zhat + \sqrt{\kappa_{a,\RR}} \xihat_{a,\RR}  \\
& + & \sqrt{\kappa_{a,\LL}} \left(\delta x + \im \delta y + \xihat_{a,\LL}\right) + \sqrt{\kappa_{a,\M}} \xihat_{a,\M} \notag \\
\dot{\dhat}_b & = & - \left(\frac{\kappa_b}{2} - \im \Delta_b \right) \dhat_b + \im \beta \zhat + \sqrt{\kappa_{b,\RR}} \xihat_{b,\RR} \notag \\
& + & \sqrt{\kappa_{b,\LL}} \left[\frac{\bar{b}_0}{\bar{a}_0} \left(\delta x + \im \delta y \right) + \xihat_{b,\LL}\right] + \sqrt{\kappa_{b,\M}} \xihat_{b,\M} \nonumber \\
  \dot{\cde} & = & - \left(\frac{\gamma}{2} + \im \omega_\M\right)\cde  - \im \left(\alpha^\ast \dhat_a + \alpha \dhat_a^\dagger - \beta^\ast \dhat_b - \beta \dhat^\dagger_b\right) \notag \\
& + & \sqrt{\gamma} \etahat , \nonumber
\end{eqnarray}
with $\alpha = A\bar{a}$ and $\beta = B \bar{b}$. The mean cavity amplitudes are 
\begin{eqnarray}
  \label{eq:meanAmplitudeTwoModes}
  \bar{a} & = & \frac{\sqrt{\kappa_{a,\LL}}}{\kappa_{a,\LL}/2 - \im \Delta_a} \bar{a}_0 \\
  \bar{b} & = & \frac{\sqrt{\kappa_{b,\LL}}}{\kappa_{b,\LL}/2 - \im \Delta_b} \bar{b}_0 . \nonumber
\end{eqnarray}
The condition stated above for a zero mean radiation pressure on the oscillator is $A|\bar{a}|^2 = B|\bar{b}|^2$.

The solutions to the equations of motion are given in Appendix \ref{app:TwoModes}, as well as exact results for the cross-correlation (\ref{eq:Sdef}) in the case of two optical modes. An interesting feature is that when, in addition to $A|\bar{a}|^2 = B|\bar{b}|^2$, $\kappa_{a,i} = \kappa_{b,i}$, $\Delta_a = \Delta_b$ are satisfied, the position operator $\zhat$ becomes independent of the classical noise $\delta x$ and $\delta y$, such that $S_\mathrm{cl,z}[\omega]$ vanishes. The properties of the quantum and thermal contributions will be the same as discussed in Sections \ref{sec:detection} and \ref{sec:quanttherm}. Although this requires fine-tuning, one might be able to get close to these conditions and thereby reduce the classical contribution significantly. 

\section{Concluding remarks}
\label{sec:conclusion}

We have presented a detailed theoretical study of a proposed experiment designed to observe radiation pressure shot noise by detecting a correlation between photon shot noise in an optical cavity mode and position of a mechanical oscillator. The experiment involves the measurement of a cross-correlator of two outgoing optical field quadratures from the cavity. We have investigated the possibility of detecting radiation pressure shot noise by this method.

The cross-correlation measurement has contributions from three noise sources: radiation pressure shot noise, classical radiation pressure noise and thermal noise in the mechanical oscillator. We have determined how the radiation pressure shot noise contribution differs qualitatively and quantitatively from the other contributions to this cross-correlator. As pointed out in Ref.~\cite{Heidmann1997ApplPhysB}, the contribution from thermal noise disappears when the drive frequency equals the cavity resonance frequency. We found that in the general case of nonzero detuning, it also disappears at a specific choice of quadratures for the cross-correlation measurement. We have presented figures of merit for when the shot noise contribution can be expected to dominate over the thermal noise contribution in the vicinity of these idealized situations. The choice of parameters in the examples we presented is relevant to present-day experiments \cite{Thompson2008Nature,Jayich2008NJP}, such that the observation of radiation pressure shot noise by this method should be within reach. 

The relative importance of classical laser noise versus quantum shot noise has also been investigated. We find that this can be diminished by the use of an asymmetric cavity and/or a cavity of very high finesse. In addition, we proposed a new setup for the membrane-in-the-middle geometry involving two optical modes. This setup can overcome the problem of static bistability, and can potentially also reduce the unwanted classical radiation pressure noise significantly.

\begin{acknowledgments}
We thank Nathan Flowers-Jacobs for valuable insights. KB acknowledges financial support from the Research Council of Norway, Grant No.~191576/V30 (FRINAT). We acknowledge support from NSF, Grant No.~0603369 (AN, SMG), 0653377 (AN, SMG, JGEH), and 0855455 (BMZ, CY, JGEH). BMZ, CY and JGEH also acknowledge support from AFOSR, Grant No.~FA9550-90-1-0484. This material is based upon work supported by DARPA under award No.~N66001-09-1-2100.
\end{acknowledgments}

\appendix
\section{Mathematical details in the case of one optical mode}
 \label{app:genExp}


We begin by writing down the solution to the equations of motion (\ref{eq:linearEOM}) in Fourier space \footnotemark[35]: 
\begin{eqnarray}
\label{eq:zSolution} 
\zhat[\omega] & = & \frac{1}{N[\omega]} \Big[ \sqrt{\gamma} \left(\chi_\M^{-1 \, \ast}[-\omega] \etahat[\omega]  + \chi_\M^{-1}[\omega] \etahat^\dagger[\omega]\right)  \\ 
 & & \qquad -  2\omega_\M \left(\alpha^\ast \chi_\C[\omega] \zetahat[\omega] + \alpha \chi^\ast_\C[-\omega]  \zetahat^\dagger[\omega] \right) \Big]  \nonumber \\
\dhat[\omega] & = & -\chi_\C[\omega] \left(\im \alpha \zhat[\omega] - \zetahat[\omega] \right) \nonumber .
\end{eqnarray}
We have defined the operator
\begin{eqnarray}
\zetahat[\omega] & = & \sqrt{\kappa_\LL} \left( \delta x[\omega] + \im \delta y[\omega] + \xihat_\LL[\omega] \right) \\
& + & \sqrt{\kappa_\RR} \xihat_\RR[\omega] + \sqrt{\kappa_\M} \xihat_\M[\omega] \ , \nonumber 
\end{eqnarray}
the susceptibilites $\chi_\C[\omega]  =  [\kappa/2 - \im(\omega + \Delta)]^{-1}$ and $\chi_\M[\omega]  = [\gamma/2 - \im(\omega - \omega_\M) ]^{-1}$, and the functions 
\begin{eqnarray}
N[\omega] & = & \chi^{-1}_\M[\omega]  \chi^{-1 \, \ast}_\M[-\omega]  + 2 \omega_\M \Sigma[\omega] , \\
\Sigma[\omega] & = & -\im |\alpha|^2 \left(\chi_\C[\omega] - \chi_\C^\ast[-\omega] \right) .  \nonumber
\end{eqnarray}
In the weak coupling limit, $-2 \im \omega_\M N^{-1}[\omega]$ can be thought of as an effective mechanical susceptibility and $\Sigma[\omega]$ as the optomechanical self-energy. 

It is instructive to write out some of the terms in Equation (\ref{eq:dXdYcont}). We begin with the (q,z)-term,
\begin{eqnarray}
    & & \langle \delta \Xhat[\omega] \delta \Yhat_\theta[\omega'] \rangle_{\mathrm{q,z}}  =  -|\alpha|^{-1} \sqrt{\kappa_\RR} \ex^{\im \theta} \Sigma[\omega] \\ 
& & \times \left(\langle \zhat[\omega] \xihat^\dagger_\LL[\omega'] \rangle - \sqrt{\kappa_\LL} \chi_\C^\ast[-\omega'] \sum_i \sqrt{\kappa_i} \langle \zhat[\omega] \xihat^\dagger_i[\omega'] \rangle \right) \nonumber \\
&  & - |\alpha|^{-1} \sqrt{\kappa_\LL} \ex^{- \im \phi} \Lambda[\omega'] \notag \\ 
& & \left( \langle \xihat_\RR[\omega] \zhat[\omega'] \rangle - \sqrt{\kappa_\RR} \chi_\C[\omega] \sum_i\sqrt{\kappa_i} \langle \xihat_i[\omega] \zhat[\omega'] \rangle \right) \ . \nonumber
\end{eqnarray}
The summations are over $i = \mathrm{L,R,M}$. We see that this term consists of correlations we wish to detect, namely correlations between the quantum vacuum noise of the electromagnetic field and position of the mechanical oscillator. We have defined the function
\begin{equation}
  \label{eq:Lambdadef}
  \Lambda[\omega] = -\im |\alpha|^2 \left( \ex^{\im (\phi-\theta)} \chi_\C[\omega] - \ex^{-\im (\phi-\theta)} \chi^\ast_\C[-\omega] \right) ,
\end{equation}
where $\phi$ is the complex phase of the mean cavity amplitude $\bar{a}$.
The term $\langle \delta \Xhat[\omega] \delta \Yhat_\theta[\omega'] \rangle_{\mathrm{cl,z}}$ contains correlation functions of the kind $\langle \delta x[\omega] \zhat[\omega'] \rangle$, whereas the last term in Equation (\ref{eq:dXdYcont}) is
\begin{eqnarray}
  & & \langle \delta \Xhat[\omega] \delta \Yhat_\theta[\omega'] \rangle_{\mathrm{z,z}} \\
& & \qquad \qquad = \sqrt{\kappa_\LL \kappa_\RR} |\alpha|^{-2} \Sigma[\omega] \Lambda[\omega'] \langle z[\omega] z[\omega'] \rangle \ \notag .  
\end{eqnarray}

We next present the terms in Equation (\ref{eq:Sparts}). The first term, which is the term of interest, is
\begin{eqnarray}
  \label{eq:Sqz}
  S_\mathrm{q,z}[\omega] & = & - \omega_\M \sqrt{\kappa_\LL \kappa_\RR} \\
 & \times & \Bigg[\frac{\Sigma[\omega]}{N[\omega]} \left(\ex^{-\im(\phi- \theta)} \chi_\C^\ast[\omega] + \ex^{\im (\phi-\theta)} \chi_\C[-\omega] \right) \notag \\
& + & \frac{\Lambda[-\omega]}{N[-\omega]} \left(\chi_\C[\omega] + \chi_\C^\ast[-\omega] \right) \Bigg] \ . \nonumber 
\end{eqnarray}
The contributions from the classical noise in the drive, when expressed by the functions
\begin{eqnarray}
  \label{eq:BDdef}
  B_\pm[\omega] & = & \ex^{-\im \phi} \chi_\C[\omega] \pm \ex^{\im \phi} \chi_\C^\ast[-\omega] \\
  D_\pm[\omega] & = & \ex^{-\im \theta} \left(1 - \kappa_\LL \chi_C[\omega] \right) \pm \ex^{\im \theta} \left(1 - \kappa_\LL \chi_\C^\ast[-\omega]\right)  \nonumber \\
 C_{B,B} [\omega]& = & \Big(B_+[\omega] B_+[-\omega] C_X - B_-[\omega] B_-[-\omega] C_Y \Big)   \nonumber\\
 C_{B,D}[\omega] & = & \Big(B_+[\omega] D_+[-\omega] C_X - B_-[\omega] D_-[-\omega] C_Y \Big),  \nonumber
\end{eqnarray}
become
\begin{eqnarray}
  \label{eq:Sclass}
  S_\mathrm{cl,cl}[\omega] & = & -\sqrt{\kappa_\LL \kappa_\RR} C_{B,D}[\omega] \\
  S_\mathrm{cl,z}[\omega] & = & 2 \omega_\M \sqrt{\kappa_\LL \kappa_\RR} \notag \\
 & \times & \Bigg( \frac{\Sigma[\omega]}{N[\omega]} C_{B,D}[\omega] - \kappa_\LL \frac{\Lambda[-\omega]}{N[-\omega]} C_{B,B}[\omega] \Bigg)  \ . \nonumber
\end{eqnarray}
The last term can be written
\begin{eqnarray}
  \label{eq:Szz}
  & & S_\mathrm{z,z}[\omega]  = \sqrt{\kappa_\LL \kappa_\RR} \frac{\Sigma[\omega] \Lambda[-\omega]}{N[\omega] N[-\omega]} \\
& & \times  \Bigg[ |\alpha|^{-2} \gamma \left(n_\mathrm{th} + \frac{1}{2}\right) \Big(|\chi_\M^{-1}[-\omega]|^2 + |\chi_\M^{-1}[\omega]|^2 \Big) \nonumber \\
& & + 4 \omega_\M^2 \kappa_\LL C_{B,B}[\omega] + 2 \omega_\M^2 \kappa \Big(|\chi_\C[\omega]|^2 + |\chi_\C[-\omega]|^2\Big) \Bigg] \nonumber  \ .
\end{eqnarray}
The first term in $S_\mathrm{z,z}[\omega]$ is typically the dominant one, which represents the fluctuations of the oscillator due to thermal noise from the mechanical bath. 

As stated earlier, in the limit $\tilde{\gamma}, |\omega - \omega_\M| \ll \kappa, \omega_\M$, the above expressions simplify. The coefficients appearing in Equation (\ref{eq:SzzSimplified}) are
\begin{eqnarray}
  \label{eq:Kthdef}
 R^\mathrm{(th)}_1 & = & K^\mathrm{(th)} \left[\left(\left(\frac{\kappa}{2}\right)^2 - \Delta^2 \right) \sin \theta - \kappa \Delta \cos \theta \right] \notag \\
I^\mathrm{(th)}_1 & = & K^\mathrm{(th)} \omega_\M \left( \frac{\kappa}{2} \sin \theta - \Delta \cos \theta \right) 
\end{eqnarray}
where
\begin{equation}
  \label{eq:K0thdef}
  K^\mathrm{(th)}  =  -2 \Delta \tilde{\gamma} \left(n_\M+\frac{1}{2}\right) K^\mathrm{(q)} \ ,
\end{equation}
and 
\begin{equation}
  \label{eq:K0qdef}
  K^\mathrm{(q)}  =  \frac{2 \sqrt{\kappa_\LL \kappa_\RR} |\alpha|^2 |\chi_\C[\omega_\M]|^2 |\chi_\C[-\omega_\M]|^2}{\sqrt{\left(\kappa/2\right)^2 + \Delta^2}}  \ .
\end{equation}
The effective phonon number $n_\M$ is given by Equation (\ref{eq:nMeff}), with
\begin{equation}
  \label{eq:noptdef}
  n_\mathrm{opt} = - \frac{\kappa |\chi_\C[-\omega_\M]|^2 + \kappa_\LL C_{B,B}[-\omega_\M]}{4 \Delta \kappa \omega_\M |\chi_\C[\omega_\M]|^2|\chi_\C[-\omega_\M]|^2} \ .
\end{equation}
The coefficients in Equation (\ref{eq:SqzSimplified}) are
\begin{eqnarray}
 \label{eq:Kqdef}
 R^\mathrm{(q)}_1 & = & K^\mathrm{(q)} \tilde{\gamma} \omega_\M \Delta \left(\Delta \sin \theta + \frac{\kappa}{2} \cos \theta \right) \\ 
R^\mathrm{(q)}_2 & = & K^\mathrm{(q)} \Bigg[\frac{\kappa}{2} \left( 3 \Delta^2 - \left(\frac{\kappa}{2}\right)^2 - \omega_\M^2 \right) \sin \theta \notag \\
& + & \Delta \left( 3 \left(\frac{\kappa}{2}\right)^2 - \Delta^2 + \omega_\M^2 \right) \cos \theta \Bigg] \nonumber \\
I^\mathrm{(q)}_1 & = & -K^\mathrm{(q)}\frac{\tilde{\gamma}}{2} \left(\left(\frac{\kappa}{2}\right)^2+\Delta^2+\omega_\M^2 \right) \notag \\
&\times & \left( \frac{\kappa}{2} \sin \theta - \Delta \cos \theta \right) . \nonumber
\end{eqnarray}
The coefficients in Equation (\ref{eq:SclzSimplified}) are quite complicated and therefore not presented here.

\subsection{Zero detuning}
\label{sec:coeffResonance}
For $|\Delta| \ll \kappa,\omega_\M$ and $\theta \neq 0$, the coefficients in Equation (\ref{eq:SzzSimplified}) become
\begin{eqnarray}
  \label{eq:KthdefDelta0}
  R^\mathrm{(th)}_1 & = & \frac{-2 \sqrt{\kappa_\LL \kappa_\RR} \kappa \Delta |\alpha|^2 \tilde{\gamma} \left(n_\M + \frac{1}{2}\right)}{ \left[(\kappa/2)^2+\omega_\M^2\right]^2} \sin \theta 
\end{eqnarray}
and $I^\mathrm{(th)}_1/R^\mathrm{(th)}_1   =  2 \omega_\M/\kappa$. For $S_\mathrm{q,z}[\omega]$, we get
\begin{eqnarray}
  \label{eq:KqdefDelta0} 
  R^\mathrm{(q)}_2 & = & -\frac{2 \sqrt{\kappa_\LL \kappa_\RR} |\alpha|^2}{ \left[(\kappa/2)^2+\omega_\M^2\right]} \sin \theta \ ,
\end{eqnarray}
$R^\mathrm{(q)}_1/(R^\mathrm{(q)}_2 \tilde{\gamma}) =  0$ and $I^\mathrm{(q)}_1/(R^\mathrm{(q)}_2 \tilde{\gamma}) =  1/2$.
In this case, the coefficients in $S_\mathrm{cl,z}[\omega]$ are relatively simple and given by
\begin{eqnarray}
 \label{eq:KcldefDelta0}
  R^\mathrm{(cl)}_2 & = & -\frac{4 \sqrt{\kappa_\LL \kappa_\RR} \, \kappa \, \kappa_\LL |\alpha|^2 C_X \sin \theta}{ \left[(\kappa/2)^2+\omega_\M^2\right]^2} \ ,
\end{eqnarray}
$R^\mathrm{(cl)}_1/(R^\mathrm{(cl)}_2 \tilde{\gamma})  =  \omega_\M/\kappa$, $I^\mathrm{(cl)}_2/R^\mathrm{(cl)}_2  = - 2 \omega_\M/\kappa $ and
$I^\mathrm{(cl)}_1 /(R^\mathrm{(cl)}_2 \tilde{\gamma}) =  -1/2$. As expected, the oscillator only sees the amplitude noise and not the phase noise when the frequency of the drive is close to the cavity resonance frequency.

It is worth noting that $|I^\mathrm{(cl)}_2/R^\mathrm{(cl)}_2| = 2\omega_\M/\kappa$, whereas $|I^\mathrm{(q)}_1/(R^\mathrm{(q)}_2 \tilde{\gamma})| = 1/2$. Unless $\kappa \gg \omega_\M$, this means that if the classical signal is dominant in the real part of $S[\omega]$, it should also be dominant in the imaginary part (note also that $|I^\mathrm{(th)}_1/R^\mathrm{(th)}_1| = |I^\mathrm{(cl)}_2/R^\mathrm{(cl)}_2|$). Thus, if a sign change is observed in $R[\omega]$, but not in $I[\omega]$, one might be able to conclude that the sign change is due to photon shot noise in the cavity and not classical noise in the drive. Another possibility is to add classical noise deliberately, in order to determine the source of the sign change in $R[\omega]$.

\subsection{Finite detuning}

The angle $\theta_c$ was defined as the quadrature angle where the constant $R^\mathrm{(th)}_1$ vanishes for a given value of $\Delta/\kappa$. For $\delta \theta = \theta -\theta_c$ and $|\delta \theta| \ll 1$, we find \footnote{Note that there are two critical angles $\theta_c$ with a difference of $\pi$. We have chosen one of them here. Choosing the other gives an overall minus sign in all coefficients.} that $R^\mathrm{(th)}_1 = K^\mathrm{(th)} \left((\kappa/2)^2+\Delta^2\right) \delta \theta$. The imaginary part does not change significantly for $\theta$ around $\theta_c$. Its value at $\theta = \theta_c$ is $I^\mathrm{(th)}_1 = K^\mathrm{(th)} \omega_\M \Delta$.
We also quote the coefficients appearing in Equations (\ref{eq:SqzSimplified}) at $\theta = \theta_c$. They are
\begin{eqnarray}
 \label{eq:KqdefThetaCrit}
 R^\mathrm{(q)}_1 & = & K^\mathrm{(q)} \frac{\tilde{\gamma} \omega_\M \kappa \Delta}{2} \\ 
R^\mathrm{(q)}_2 & = & K^\mathrm{(q)} \Delta \left(\left(\frac{\kappa}{2}\right)^2 + \Delta^2 - \omega_\M^2\right) \nonumber \\
I^\mathrm{(q)}_1 & = & - K^\mathrm{(q)} \frac{\tilde{\gamma} \Delta}{2} \left(\left(\frac{\kappa}{2}\right)^2+\Delta^2+\omega_\M^2 \right) \ . \nonumber
\end{eqnarray}

\section{Mathematical details in the case of two optical modes}
\label{app:TwoModes}

The solution to Equations (\ref{eq:EOMTwoModes}) are
\begin{eqnarray}
\zhat[\omega] & = & \frac{1}{N_2[\omega]} \Big[ \sqrt{\gamma} \left(\chi_\M^{-1 \, \ast}[-\omega] \eta[\omega]  + \chi_\M^{-1}[\omega] \eta^\dagger[\omega]\right)  \nonumber\\ 
 & & \qquad -  2\omega_\M \Big(\alpha^\ast \chi_a[\omega]
   \zeta_a[\omega] + \alpha \chi^\ast_a[-\omega]  \zeta_a^\dagger[\omega]
   \notag \\
& & \qquad - \beta^\ast \chi_b[\omega]
   \zeta_b[\omega] - \beta \chi^\ast_b[-\omega]  \zeta_b^\dagger[\omega]
 \Big) \Big]  \\
\dhat_a[\omega] & = & -\chi_a[\omega] \left(\im \alpha \zhat[\omega] - \zeta_a[\omega] \right)  \nonumber \\
\dhat_b[\omega] & = & \chi_b[\omega] \left(\im \beta \zhat[\omega] + \zeta_b[\omega] \right) .  \nonumber
\end{eqnarray}
We have defined
\begin{eqnarray}
\zeta_a[\omega] & = & \sqrt{\kappa_{a,\LL}} \left( \delta x[\omega] + \im \delta y[\omega] + \xi_{a,\LL}[\omega] \right) \\
& + & \sqrt{\kappa_{a,\RR}} \xi_{a,\RR}[\omega] + \sqrt{\kappa_{a,\M}} \xi_{a,\M}[\omega] \notag \\
\zeta_b[\omega] & = & \sqrt{\kappa_{b,\LL}} \left[ \frac{\bar{b}_0}{\bar{a}_0}\left( \delta x[\omega] + \im \delta y[\omega]\right) + \xi_{b,\LL}[\omega] \right] \notag \\ 
& + & \sqrt{\kappa_{b,\RR}} \xi_{b,\RR}[\omega] + \sqrt{\kappa_{b,\M}} \xi_{b,\M}[\omega]   \nonumber
\end{eqnarray}
and 
\begin{eqnarray}
N_2[\omega] & = & \chi^{-1}_\M[\omega]  \chi^{-1 \, \ast}_\M[-\omega]  +
2 \omega_\M \Sigma_2[\omega]  \\
\Sigma_2[\omega] & = & \Sigma_a[\omega] + \Sigma_b[\omega]  \nonumber\\
\Sigma_a[\omega] & = & -\im |\alpha|^2 \left(\chi_a[\omega] -
  \chi_a^\ast[-\omega] \right)  \nonumber\\
\Sigma_b[\omega] & = & -\im |\beta|^2 \left(\chi_b[\omega] -
  \chi_b^\ast[-\omega] \right)  \nonumber \ ,
\end{eqnarray}
where the optical mode susceptibilities are $\chi_{a/b}[\omega]  =  \left[\kappa_{a/b}/2 - \im(\omega + \Delta_{a/b})\right]^{-1}$.
As stated in Section \ref{sec:twomodes}, the mean radiation pressure on the oscillator is zero when $A|\bar{a}|^2 = B|\bar{b}|^2$. If in addition $\kappa_{a,i} = \kappa_{b,i}$ and $\Delta_a = \Delta_b$, it is straightforward to check that $\zhat$ becomes independent of the classical noise $\delta x$ and $\delta y$. This requires fine-tuning, but it shows that it might be possible to make the classical contribution small enough to be negligible.

We now return to the general case and imagine measuring the cross-correlation between the quadrature fluctuations
\begin{equation}
  \label{eq:dXdefTwoModes}
  \delta \Xhat(t) = \ex^{-\im \phi_a} \dhat_{a,\mathrm{out},\RR}(t) + \ex^{\im \phi_a} \dhat^\dagger_{a,\mathrm{out},\RR}(t)
\end{equation}
and 
\begin{equation}
  \label{eq:dYdefTwoModes}
  \delta \Yhat_\theta(t) = \ex^{-\im \theta} \dhat_{b,\mathrm{out},\LL}(t) + \ex^{\im \theta} \dhat^\dagger_{b,\mathrm{out},\LL}(t)
\end{equation}
in the same way as before, with $\dhat_{j,\mathrm{out},i}(t) = \ex^{\im \omega_{\mathrm{D},j} t}(\hat{a}_{j,\mathrm{out},i}(t) - \langle \hat{a}_{j,\mathrm{out},i} (t) \rangle)$ and $\ex^{\im\phi_a} = \alpha/|\alpha|$. $\delta \Xhat$ represents the intensity fluctuations in the output of mode $a$ on the right hand side of the cavity. $\delta \Yhat_\theta$ is the fluctuation in an arbitrary quadrature in the output of mode $b$ on the left hand side. It is however not essential whether mode $a$ or $b$ is detected. Defining $S[\omega]$ as in Equation (\ref{eq:Sdef}), it still has four contributions as in Equation (\ref{eq:Sparts}). The quantum contribution is
\begin{eqnarray}
  \label{eq:SqzTwoModes}
  & & S_\mathrm{q,z}[\omega]  =  \omega_\M  \\
  &  & \times  \left[\sqrt{\kappa_{a,\LL} \kappa_{b,\RR}} \frac{|\beta|}{|\alpha|} \frac{\Sigma_a[\omega]}{N_2[\omega]} \left(\ex^{-\im \lambda} \chi^\ast_b[\omega] + \ex^{\im \lambda} \chi_b[-\omega] \right) \right. \nonumber \\
 & & \left. +
\sqrt{\kappa_{a,\RR} \kappa_{b,\LL}} \frac{|\alpha|}{|\beta|} \frac{\Lambda_b[-\omega]}{N_2[-\omega]} \Big(\chi_a[\omega] + \chi_a^\ast[-\omega]\Big) \right] \nonumber
\end{eqnarray}\vspace{0.1cm}\\
with $\Lambda_b[\omega] = -\im |\beta|^2 \left( \ex^{\im \lambda} \chi_b[\omega] - \ex^{-\im \lambda} \chi^\ast_b[-\omega] \right)$, $\lambda = (\phi_b - \theta)$ and $\ex^{\im \phi_b} = \beta/|\beta|$. The terms due to classical noise in the drive can be expressed by the functions
\begin{eqnarray}
  \label{eq:BDEdefTwoMode}
  B_{a,\pm}[\omega] & = & \ex^{-\im \phi_a} \chi_a[\omega] \pm \ex^{\im \phi_a} \chi^\ast_a[-\omega] \\
  B_{b,\pm}[\omega] & = & \frac{\bar{b}_0}{\bar{a}_0} \left(\ex^{-\im \phi_b} \chi_b[\omega] \pm \ex^{\im \phi_b} \chi^\ast_b[-\omega] \right)  \nonumber\\
  D_{b,\pm}[\omega] & = & \frac{\bar{b}_0}{\bar{a}_0} \Big[ \ex^{-\im \theta}\left(1 - \kappa_{b,\LL} \chi_b[\omega]\right) \notag \\
 & \pm & \ex^{\im \theta} \left(1 - \kappa_{b,\LL} \chi^\ast_b[-\omega]\right) \Big]   \nonumber\\
  E_\pm[\omega] & = & \sqrt{\kappa_{a,\LL}} |\alpha| B_{a,\pm}[\omega] - \sqrt{\kappa_{b,\LL}} |\beta| B_{b,\pm}[\omega] .  \nonumber
\end{eqnarray}
They become
\begin{eqnarray}
  \label{eq:SclTwoModes}
  S_\mathrm{cl,cl}[\omega] & = & -\sqrt{\kappa_{a,\LL} \kappa_{a,\RR}} C_{B_a, D_b}[\omega] , \\
 S_ \mathrm{cl,z}[\omega] & = & 2 \omega_\M \sqrt{\kappa_{a,\RR}} \Bigg( \frac{1}{|\alpha|} \frac{\Sigma_a[\omega]}{N_2[\omega]} C_{D_b, E}[\omega] \notag \\
& + & \sqrt{\kappa_{a,\LL} \kappa_{b,\LL}} \frac{1}{|\beta|} \frac{\Lambda_b[-\omega]}{N_2[-\omega]} C_{B_a, E}[\omega] \Bigg) \ , \notag
\end{eqnarray}
where we have used the abbreviation $C_{B_a,D_b}[\omega] = B_{a,+}[\omega] D_{b,+}[-\omega] C_X - B_{a,-}[\omega] D_{b,-}[-\omega] C_Y$ and similarly for $C_{B_a, E}[\omega]$ etc.
Finally, the last contribution is 
\begin{eqnarray}
  \label{eq:SzzTwoModes}
 & & S_\mathrm{z,z}[\omega] = -\sqrt{\kappa_{a,\RR} \kappa_{b,\LL}} \frac{1}{|\alpha||\beta|} \frac{\Sigma_a[\omega] \Lambda_b[-\omega]}{N_2[\omega]  N_2[-\omega]} \\
  &  & \times \Bigg\{\gamma \left(n_\mathrm{th} + \frac{1}{2} \right)\Big(|\chi_M^{-1}[-\omega]|^2 + |\chi_M^{-1}[\omega]|^2\Big) \nonumber \\
 & & + 4 \omega_\M^2 C_{E,E}[\omega] + 2 \omega_\M^2 \Big[\kappa_a |\alpha|^2 \left(|\chi_a[\omega]|^2 + |\chi_a[-\omega]|^2\right) \notag \\
& & + \kappa_b |\beta|^2 \left(|\chi_b[\omega]|^2 + |\chi_b[-\omega]|^2\right) \Big] \Bigg\} .  \nonumber
\end{eqnarray}
Again, we point out that when $A|\bar{a}|^2 = B|\bar{b}|^2$, $\Delta_a = \Delta_b$, $\kappa_{a,i} = \kappa_{b,i}$, the functions $E_\pm[\omega]$ are identically zero and hence $S_\mathrm{cl,z}[\omega] = 0$. In this case, when analyzing the other contributions, one finds the same results as in the case with one optical mode, only with modified $\gamma_\mathrm{opt}$ and $\delta \omega_\M$.


\end{document}